%
%
%
%
%
\catcode`@=11 
%

\font\fourteenrm=cmr10 scaled\magstep2
\font\twelverm=cmr10 scaled\magstep1
\font\elevenrm=cmr10 scaled\magstephalf
\font\ninerm=cmr9

\font\sevenrm=cmr7
\font\sixrm=cmr6
\font\fiverm=cmr5

\font\fourteenbf=cmbx10 scaled\magstep2
\font\twelvebf=cmbx10 scaled\magstep1
\font\elevenbf=cmbx10 scaled\magstephalf
\font\ninebf=cmbx9
\font\sixbf=cmbx6
\font\twentyi=cmmi12 scaled\magstep3	   \skewchar\twentyi='177
\font\seventeeni=cmmi12 scaled\magstep2	   \skewchar\seventeeni='177
\font\fourteeni=cmmi10 scaled\magstep2	   \skewchar\fourteeni='177
\font\twelvei=cmmi10 scaled\magstep1	   \skewchar\twelvei='177
\font\eleveni=cmmi10 scaled\magstephalf    \skewchar\eleveni='177
\font\ninei=cmmi9			   \skewchar\ninei='177
\font\sixi=cmmi6			   \skewchar\sixi='177
\font\twentysy=cmsy10 scaled\magstep4      \skewchar\twentysy='60
\font\seventeensy=cmsy10 scaled\magstep3   \skewchar\seventeensy='60
\font\fourteensy=cmsy10 scaled\magstep2	   \skewchar\fourteensy='60
\font\twelvesy=cmsy10 scaled\magstep1	   \skewchar\twelvesy='60
\font\elevensy=cmsy10 scaled\magstephalf   \skewchar\elevensy='60
\font\ninesy=cmsy9			   \skewchar\ninesy='60
\font\sixsy=cmsy6			   \skewchar\sixsy='60

\font\fourteenex=cmex10 scaled\magstep2
\font\twelveex=cmex10 scaled\magstep1
\font\elevenex=cmex10 scaled\magstephalf

\font\fourteensl=cmsl10 scaled\magstep2
\font\twelvesl=cmsl10 scaled\magstep1
\font\elevensl=cmsl10 scaled\magstephalf
\font\ninesl=cmsl9

\font\fourteenit=cmti10 scaled\magstep2
\font\twelveit=cmti10 scaled\magstep1
\font\elevenit=cmti10 scaled\magstephalf
\font\twelvett=cmtt10 scaled\magstep1
\font\eleventt=cmtt10 scaled\magstephalf

\font\twelvecp=cmcsc10 scaled\magstep1
\font\elevencp=cmcsc10 scaled\magstephalf
\font\tencp=cmcsc10
\newfam\cpfam
\newcount\f@ntkey \f@ntkey=0
\def\samef@nt{\relax \ifcase\f@ntkey \rm \or\oldstyle \or\or
	 \or\it \or\sl \or\bf \or\tt \or\caps \fi}
\def\fourteenpoint{\relax%
        \textfont0=\fourteenrm \scriptfont0=\tenrm%
        \scriptscriptfont0=\sevenrm%
    \def\rm{\fam0 \fourteenrm \f@ntkey=0 }\relax%
        \textfont1=\fourteeni \scriptfont1=\teni
        \scriptscriptfont1=\seveni%
    \def\oldstyle{\fam1 \fourteeni\f@ntkey=1 }\relax%
        \textfont2=\fourteensy \scriptfont2=\tensy%
        \scriptscriptfont2=\sevensy%
        \textfont3=\fourteenex \scriptfont3=\fourteenex%
        \scriptscriptfont3=\fourteenex%
    \def\it{\fam\itfam \fourteenit\f@ntkey=4 }%
        \textfont\itfam=\fourteenit%
    \def\sl{\fam\slfam \fourteensl\f@ntkey=5 }%
        \textfont\slfam=\fourteensl \scriptfont\slfam=\tensl%
    \def\bf{\fam\bffam \fourteenbf\f@ntkey=6 }
        \textfont\bffam=\fourteenbf
        \scriptfont\bffam=\tenbf \scriptscriptfont\bffam=\sevenbf
    \def\tt{\fam\ttfam \twelvett \f@ntkey=7 }%
        \textfont\ttfam=\twelvett%
        \h@big=11.9\p@{} \h@Big=16.1\p@{}%
        \h@bigg=20.3\p@{} \h@Bigg=24.5\p@{}%
    \def\caps{\fam\cpfam \twelvecp \f@ntkey=8 }%
        \textfont\cpfam=\twelvecp%
    \setbox\strutbox=\hbox{\vrule height12pt depth5pt width\z@ }%
    \samef@nt }
\def\twelvepoint{\relax
        \textfont0=\twelverm \scriptfont0=\ninerm
        \scriptscriptfont0=\sixrm
    \def\rm{\fam0 \twelverm \f@ntkey=0 }\relax
        \textfont1=\twelvei \scriptfont1=\ninei
        \scriptscriptfont1=\sixi
    \def\oldstyle{\fam1 \twelvei\f@ntkey=1 }\relax
        \textfont2=\twelvesy \scriptfont2=\ninesy
        \scriptscriptfont2=\sixsy
        \textfont3=\twelveex \scriptfont3=\twelveex
        \scriptscriptfont3=\twelveex
    \def\it{\fam\itfam \twelveit \f@ntkey=4 }
        \textfont\itfam=\twelveit
    \def\sl{\fam\slfam \twelvesl \f@ntkey=5 }
        \textfont\slfam=\twelvesl \scriptfont\slfam=\ninesl
    \def\bf{\fam\bffam \twelvebf \f@ntkey=6 \relax}
        \textfont\bffam=\twelvebf \scriptfont\bffam=\ninebf
        \scriptscriptfont\bffam=\sixbf
    \def\tt{\fam\ttfam \twelvett \f@ntkey=7 }
        \textfont\ttfam=\twelvett
        \h@big=10.2\p@{} \h@Big=13.8\p@{}
        \h@bigg=17.4\p@{} \h@Bigg=21.0\p@{}
    \def\caps{\fam\cpfam \twelvecp \f@ntkey=8 }
        \textfont\cpfam=\twelvecp
    \setbox\strutbox=\hbox{\vrule height 10pt depth 4pt width\z@ }
    \samef@nt }
\def\elevenpoint{\relax
        \textfont0=\elevenrm \scriptfont0=\sevenrm
        \scriptscriptfont0=\fiverm
    \def\rm{\fam0 \elevenrm \f@ntkey=0 }\relax
        \textfont1=\eleveni \scriptfont1=\seveni
        \scriptscriptfont1=\fivei
    \def\oldstyle{\fam1 \eleveni \f@ntkey=1 }\relax
        \textfont2=\elevensy \scriptfont2=\sevensy
        \scriptscriptfont2=\fivesy
        \textfont3=\elevenex \scriptfont3=\elevenex
        \scriptscriptfont3=\elevenex
   \def\it{\fam\itfam \elevenit \f@ntkey=4 }\textfont\itfam=\elevenit
   \def\sl{\fam\slfam \elevensl \f@ntkey=5 }\textfont\slfam=\elevensl
   \def\bf{\fam\bffam \elevenbf \f@ntkey=6 }
        \textfont\bffam=\elevenbf \scriptfont\bffam=\sevenbf
        \scriptscriptfont\bffam=\fivebf
 \def\tt{\fam\ttfam \eleventt \f@ntkey=7 }\textfont\ttfam=\eleventt
 \def\caps{\fam\cpfam \elevencp \f@ntkey=8 }\textfont\cpfam=\elevencp
    \setbox\strutbox=\hbox{\vrule height 9.3pt depth 3.8pt width\z@ }
    \samef@nt }
\def\tenpoint{\relax
        \textfont0=\tenrm \scriptfont0=\sevenrm
        \scriptscriptfont0=\fiverm
    \def\rm{\fam0 \tenrm \f@ntkey=0 }\relax
        \textfont1=\teni \scriptfont1=\seveni
        \scriptscriptfont1=\fivei
    \def\oldstyle{\fam1 \teni \f@ntkey=1 }\relax
        \textfont2=\tensy \scriptfont2=\sevensy
        \scriptscriptfont2=\fivesy
        \textfont3=\tenex \scriptfont3=\tenex
        \scriptscriptfont3=\tenex
    \def\it{\fam\itfam \tenit \f@ntkey=4 }\textfont\itfam=\tenit
    \def\sl{\fam\slfam \tensl \f@ntkey=5 }\textfont\slfam=\tensl
    \def\bf{\fam\bffam \tenbf \f@ntkey=6 }
        \textfont\bffam=\tenbf \scriptfont\bffam=\sevenbf
        \scriptscriptfont\bffam=\fivebf
    \def\tt{\fam\ttfam \tentt \f@ntkey=7 }\textfont\ttfam=\tentt
    \def\caps{\fam\cpfam \tencp \f@ntkey=8 }\textfont\cpfam=\tencp
    \setbox\strutbox=\hbox{\vrule height 8.5pt depth 3.5pt width\z@ }
    \samef@nt }
\newdimen\h@big  \h@big=8.5\p@
\newdimen\h@Big  \h@Big=11.5\p@
\newdimen\h@bigg  \h@bigg=14.5\p@
\newdimen\h@Bigg  \h@Bigg=17.5\p@
\def\big#1{{\hbox{$\left#1\vbox to\h@big{}\right.\n@space$}}}
\def\Big#1{{\hbox{$\left#1\vbox to\h@Big{}\right.\n@space$}}}
\def\bigg#1{{\hbox{$\left#1\vbox to\h@bigg{}\right.\n@space$}}}
\def\Bigg#1{{\hbox{$\left#1\vbox to\h@Bigg{}\right.\n@space$}}}
\def\blank{ }
%

\font\twelvebi=cmmib10 scaled\magstep1
\font\elevenbi=cmmib10 
\font\tenbi=cmmib10
\newfam\bifam \def\bi{\fam\bifam\tenbi }
      \textfont\bifam=\twelvebi \scriptfont\bifam=\elevenbi
      \scriptscriptfont\bifam=\tenbi
\def\mib{\bi}                       
%
%
\tolerance=1600
\hfuzz=1pt \vfuzz=12pt
\parindent=22pt
\def\myhyphen{\tolerance=2000 \pretolerance=3600
       \hbadness=3600 \hyphenpenalty=500 \exhyphenpenalty=500
       \interlinepenalty=1000 \predisplaypenalty=9000
       \postdisplaypenalty=500 \binoppenalty=700 \relpenalty=500}
\newif\ifsingl@
\newif\ifdoubl@
\def\singlespace{\singl@true\doubl@false\spaces@t}
\def\doublespace{\singl@false\doubl@true\spaces@t}
\def\normalspace{\singl@false\doubl@false\spaces@t}
\newcount\fontsize \fontsize=12
\newif\iftwelv@ \twelv@true
\def\Tenpoint{\tenpoint \fontsize=10 \spaces@t}
\def\Elevenpoint{\elevenpoint \fontsize=11 \spaces@t}
\def\Twelvepoint{\twelvepoint \fontsize=12 \spaces@t}
\def\spaces@t{\relax%
    \ifnum\fontsize=12%
       \ifsingl@\subspaces@t3:4;\else\subspaces@t7:8;\fi\fi%
    \ifnum\fontsize=11%
       \ifsingl@\subspaces@t5:7;\else\subspaces@t6:7;\fi\fi%
    \ifnum\fontsize=10%
       \ifsingl@\subspaces@t3:5;\else\subspaces@t4:5;\fi\fi%
    \ifdoubl@ \multiply\baselineskip by 5%
              \divide\baselineskip by 4 \fi}
\def\subspaces@t#1:#2;{%
    \baselineskip=\normalbaselineskip%
    \multiply\baselineskip by #1 \divide\baselineskip by #2%
    \lineskip=\normallineskip%
    \multiply\lineskip by #1 \divide\lineskip by #2%
    \lineskiplimit=\normallineskiplimit%
    \multiply\lineskiplimit by #1 \divide\lineskiplimit by #2%
    \parskip=\normalparskip%
    \multiply\parskip by #1 \divide\parskip by #2%
    \abovedisplayskip=\normdisplayskip%
    \multiply\abovedisplayskip by #1 \divide\abovedisplayskip by #2%
    \belowdisplayskip=\abovedisplayskip%
    \abovedisplayshortskip=\normaldispshortskip%
    \multiply\abovedisplayshortskip by #1%
    \divide\abovedisplayshortskip by #2%
    \belowdisplayshortskip=\abovedisplayshortskip%
    \advance\belowdisplayshortskip by \belowdisplayskip%
    \divide\belowdisplayshortskip by 2%
    \smallskipamount=\skipamount%
    \multiply\smallskipamount by #1 \divide\smallskipamount by #2%
    \medskipamount=\smallskipamount \multiply\medskipamount by 2%
    \bigskipamount=\smallskipamount \multiply\bigskipamount by 4}

\def\figbaselines{\baselineskip=10.0pt  \lineskip=1.0pt
                  \lineskiplimit=1.0pt}
\def\normalbaselines{\baselineskip=\normalbaselineskip%
    \lineskip=\normallineskip \lineskiplimit=\normallineskip%
    \ifnum\fontsize=10
       \multiply\baselineskip by 4 \divide\baselineskip by 5%
       \multiply\lineskiplimit by 4 \divide\lineskiplimit by 5%
       \multiply\lineskip by 4 \divide\lineskip by 5 \fi
   \ifnum\fontsize=11
       \multiply\baselineskip by 6 \divide\baselineskip by 7%
       \multiply\lineskiplimit by 6 \divide\lineskiplimit by 7%
       \multiply\lineskip by 6 \divide\lineskip by 7 \fi }
\def\spacecheck#1{\dimen@=\pagegoal\advance\dimen@ by -\pagetotal
    \ifdim\dimen@<#1 \ifdim\dimen@>0pt \vfil\break \fi\fi}
\normalbaselineskip=20pt plus 0.2pt minus 0.1pt
\normallineskip=1.5pt plus 0.1pt minus 0.1pt
\normallineskiplimit=1.5pt
\newskip\normdisplayskip    \normdisplayskip=15pt plus 4pt minus 8pt
\newskip\normaldispshortskip \normaldispshortskip=6pt plus 5pt
\newskip\normalparskip      \normalparskip=2pt plus 0.2pt minus 0.2pt
\newskip\skipamount          \skipamount=5pt plus 2pt minus 1.5pt
\newskip\headskipamount	     \headskipamount=0pt plus 3pt minus 3pt
\newskip\frontpageskip       \frontpageskip=0pt plus .1fil
\def\oneskip{\vskip\baselineskip}
\def\headskip{\vskip\headskipamount}
\newdimen\normalhsize
\newdimen\halfhsize
\newdimen\normalvsize
\newdimen\halfvsize     \halfvsize=115mm
%
%
\def\backup{\penalty 100%
      \splittopskip=0pt \splitmaxdepth=\maxdimen \floatingpenalty=0%
       \begingroup \global\setbox0=\vbox{\unvbox255}%
            \dimen@=\ht0 \advance\dimen@ by\dp0%
            \advance\dimen@ by12pt \advance\dimen@ by\pagetotal%
            \ifdim\dimen@>\pagegoal%
            \p@gefalse \fi%
            \endgroup}
\newif\ifretry  \retryfalse
\def\plainoutput{\ifretry\global\retryfalse%
               \global\advance\vsize by\ht0%
         \shipout\vbox{\makeheadline\unvbox255\unvbox0\makefootline}%
         \else\shipout\vbox{\pageall}\fi%
              \advancepageno%
              \ifnum\outputpenalty>-20000\else\dosupereject\fi}
\def\pageall{\makeheadline%
             \vbox to\vsize{\boxmaxdepth=\maxdepth \pagecontents}%
             \makefootline}
\newcount\hangafterA
\newskip\hangindentA
\newcount\prevgrafA
\def\par{\hangafterA=\the\hangafter \hangindentA=\the\hangindent%
   \endgraf%
   \ifnum\prevgraf<-\hangafterA \prevgrafA=\the\prevgraf%
   \hangafter=\the\hangafterA \hangindent=\the\hangindentA%
   \indent\prevgraf=\the\prevgrafA\fi}%

\def\nextline{\ifnum\referencecount<-1 \endgraf \hskip7.0ex
              \else \unskip\nobreak\hskip\parfillskip\break \fi}
\countdef\pagenumber=1  \pagenumber=1
\newcount\chapternumber \chapternumber=0
\newcount\sectionnumber	\sectionnumber=0
\newcount\subsectionnumber \subsectionnumber=0
\newcount\equanumber \equanumber=0

\newif\iffrontpage
\def\FRONTPAGE{\ifvoid255\else\bigskip\vskip\frontpageskip
               \penalty-2000\fi
               \global\frontpagetrue \global\lastf@@t=0
               \global\footsymbolcount=0}
\newif\ifp@genum \p@genumtrue
\def\nopagenumbers{\p@genumfalse}
\def\pagenumbers{\p@genumtrue}
\def\pagecontents{%
    \ifvoid\topins\else\unvbox\topins\vskip\skip\topins\fi
    \dimen@=\dp255 \unvbox255
    \ifvoid\footins\else
       \ifnum\footsymbolcount<0\else\vskip\skip\footins \footrule\fi
       \unvbox\footins
    \fi%
    \ifr@ggedbottom \kern-\dimen@ \vfil \fi}
\def\advancepageno{\global\advance\pageno by 1
    \ifnum\pagenumber<0 \global\advance\pagenumber by -1
    \else\global\advance\pagenumber by 1 \fi
    \global\frontpagefalse}
\def\folio{\ifnum\pagenumber<0 \romannumeral-\pagenumber
	   \else \number\pagenumber \fi}
\def\fullline{\hbox to\normalhsize}
\newtoks\oddheadline  \oddheadline={ }
\newtoks\evenheadline \evenheadline={ }
\newtoks\paperheadline
\paperheadline={\hss\iffrontpage\else
             \ifodd\pageno\the\oddheadline \else\the\evenheadline \fi
             \hss \fi}
\headline={\the\paperheadline}
\def\makeheadline{\vbox to0pt{\skip@=\topskip
    \advance\skip@ by -12pt \advance\skip@ by -2\normalbaselineskip
    \vskip\skip@ \fullline{\vbox to12pt{}\the\headline}\vss}
    \nointerlineskip}
%
%
%
\def\attach#1{\space@ver{\strut^{\mkern 2mu #1}}\@sf}
%
%
\def\footnote#1{\attach{#1}\vfootnote{#1}}

\def\footrule{\dimen@=\prevdepth \nointerlineskip
    \vbox to0pt{\vskip-0.25\baselineskip \hrule width0.35\hsize \vss}
    \prevdepth=\dimen@ }
\newtoks\foottokens     \foottokens={\tenpoint\singlespace}
\newdimen\footindent    \footindent=24pt
\def\vfootnote#1{\insert\footins\bgroup  \the\foottokens
    \interlinepenalty=5000 \floatingpenalty=20000
    \splittopskip=\ht\strutbox \splitmaxdepth=\dp\strutbox
    \leftskip=\footindent \rightskip=\z@skip
    \parindent=0.5\footindent \parfillskip=0pt plus 1fil
    \spaceskip=\z@skip \xspaceskip=\z@skip
    \textindent{$ #1 $}\footstrut\futurelet\next\fo@t}
\newcount\lastf@@t	     \lastf@@t=-1
\newcount\footsymbolcount    \footsymbolcount=0
\def\fd@f#1{\xdef\footsymbol{#1}}
\let\footsymbol=\star
\def\footsymbolgen{\relax \NPsymbolgen
    \global\lastf@@t=\pageno}
\def\NPsymbolgen{%
    \ifnum\footsymbolcount<0 \global\footsymbolcount=0\fi
    {\iffrontpage \else \advance\lastf@@t by 1 \fi
    \ifnum\lastf@@t<\pageno \global\footsymbolcount=0
    \else \global\advance\footsymbolcount by 1 \fi }
    \ifcase\footsymbolcount
    \fd@f\star\or \fd@f\dagger\or \fd@f\ast\or
    \fd@f\ddagger\or \fd@f\natural\or \fd@f\diamond\or
    \fd@f\bullet\or \fd@f\nabla\else
    \fd@f\dagger\global\footsymbolcount=0 \fi}
%
%
\newcount\figurecount	  \figurecount=0
\def\fignum#1{\global\advance\figurecount by 1%
    \xdef#1{\the\figurecount}\message{[Fig.\the\figurecount]}}

\def\Figleft/#1/#2{\fignum\?\expandafter\figleft/\?/#1/{#2}}
\def\Figright/#1/#2{\fignum\?\expandafter\figright/\?/#1/{#2}}
\def\Figcenter/#1/#2{\fignum\?\expandafter\figcenter/\?/#1/{#2}}
\def\Figtop/#1/#2{\fignum\?\expandafter\figtop/\?/#1/{#2}}
\def\Figfoot/#1/#2{\fignum\?\expandafter\figfoot/\?/#1/{#2}}
\def\Figpage/#1{\fignum\?\expandafter\figpage/\?/{#1}}
\def\Figscenter/#1/#2#3{\fignum\?\fignum\!%
                       \figscenter/\?\!/#1/{#2}{#3}}
\def\Figspage/#1#2{\fignum\?\fignum\!\figspage/\?\!/{#1}{#2}}
\def\Figstop/#1/#2#3{\fignum\?\fignum\!\figstop/\?\!/#1/{#2}{#3}}
\def\Figsfoot/#1/#2#3{\fignum\?\fignum\!\figsfoot/\?\!/#1/{#2}{#3}}
\def\FIGleft#1/#2/#3{\fignum#1\figleft/#1/#2/{#3}}
\def\FIGright#1/#2/#3{\fignum#1\figright/#1/#2/{#3}}
\def\FIGcenter#1/#2/#3{\fignum#1\figcenter/#1/#2/{#3}}
\def\FIGtop#1/#2/#3{\fignum#1\figtop/#1/#2/{#3}}
\def\FIGfoot#1/#2/#3{\fignum#1\figfoot/#1/#2/{#3}}
\def\FIGpage#1/#2{\fignum#1\figpage/#1/{#2}}
\def\FIGScenter#1#2/#3/#4#5{\fignum#1\fignum#2%
                       \figscenter/#1#2/#3/{#4}{#5}}
\def\FIGSpage#1#2/#3#4{\fignum#1\fignum#2\figspage/#1#2/{#3}{#4}}
\def\FIGStop#1#2/#3/#4#5{\fignum#1\fignum#2\figstop/#1#2/#3/{#4}{#5}}
\def\FIGSfoot#1#2/#3/#4#5{\fignum#1\fignum#2%
                       \figsfoot/#1#2/#3/{#4}{#5}}
%
%
\catcode`@=11 
%
%
%
\newdimen\shiftwindow \shiftwindow=0pt
\newcount\hhfigcap
\newcount\mmfigcap
\newdimen\figcapht
\newdimen\figcapsft
\newif\if@mid
\newif\ifp@ge
\def\figcaptxt/#1/#2{\tenpoint\singlespace\figbaselines
                      Fig.#1.\ #2 \endgraf\oneskip}
\def\figcapbox/#1/#2/#3{\vbox{\endgraf%
              \parskip=0mm \parindent=0pt
              \if#1C \hangia=0.054\hsize \hangic=0.092\hsize%
                     \hangib=0.892\hsize \hangid=0.854\hsize%
              \else  \hangib=0.414\hsize \hangid=\hangib%
                     \advance\hangib by 6mm%
                     \hangia=2mm \hangic=8mm
                \if#1R \advance\hangia by 0.53\hsize
                       \advance\hangic by 0.53\hsize \fi \fi
              \parshape=2 \hangia \hangib \hangic \hangid
             {\figcaptxt/#2/{#3}}
              }}
\def\graphfile{\blank}
\def\graphcontent{\if\graphfile\blank\vss%
                  \else\vss\input\graphfile\vss\fi}
\def\figcolumn/#1/#2/#3/#4{\myhyphen%
              \FIGPAR#1{#3}%
              \global\figcapht=\baselineskip%
              \ifnum\mmfigcap>0%
                \global\multiply\figcapht by\mmfigcap%
              \else%
                \global\figcapsft=\baselineskip%
                \global\multiply\figcapsft by\mmfigcap%
                \vskip\figcapsft%
                \global\multiply\figcapht by#3%
              \fi%
              \vbox to\figcapht{\if#1R\shiftwindow=0.42\hsize\fi%
                         \graphcontent\figcapbox/#1/#2/{#4}}%
              \vskip-\figcapht%
              \ifnum\mmfigcap>0\else\vskip-\figcapsft\fi%
              \let\graphfile=\blank \shiftwindow=0pt}%
\newcount\figheadstart  \figheadstart=3
\newcount\hangbefore
\def\FIGPAR#1#2{%
    \hangid=\hsize \divide\hangid by 2 \if#1R \hangid=-\hangid \fi%
    \edef\next{\hangafter=\figheadstart \hangindent=\hangid}%
    \ifnum\figheadstart>0 \next \fi%
    \endgraf\hangbefore=\the\prevgraf \parskip=0pt \next%
    \ifnum\hangbefore>\hangafter%
      \advance\hangbefore by-\figheadstart
    \else \hangbefore=0 \penalty-3000 \fi%
    \hangafter=-#2 \advance\hangafter by\hangbefore%
    \message{(\the\figheadstart,\the\hangafter,\the\hangbefore)}%
    \ifnum\hangafter>-1 \hangafter=999 \global\mmfigcap=-#2%
    \else \global\mmfigcap=-\the\hangafter \fi}
\newskip\hangia
\newskip\hangib
\newskip\hangic
\newskip\hangid
\newskip\hfigcap
\newskip\mfigcap

\def\figtop/#1/#2/#3{%
            \@midfalse\p@gefalse\begingroup%
            \mmfigcap=#2 \global\figcapht=\normalbaselineskip%
            \global\multiply\figcapht by\mmfigcap%
            \setbox0=\vbox%
              to\figcapht{\graphcontent\figcapbox/C/#1/{#3}}%
            \insert\topins{\penalty100 \floatingpenalty=0%
            \splittopskip=0pt \splitmaxdepth=\maxdimen%
            \box0\break}\endgroup\let\graphfile=\blank}
\def\figfoot/#1/#2/#3{%
            \@midfalse\p@gefalse\begingroup%
            \mmfigcap=#2 \global\figcapht=\normalbaselineskip%
            \global\multiply\figcapht by\mmfigcap%
            \setbox0=\vbox%
              to\figcapht{\graphcontent\figcapbox/C/#1/{#3}}%
            \global\footsymbolcount=-1%
            \insert\footins{\penalty100 \floatingpenalty=0%
            \splittopskip=0pt \splitmaxdepth=\maxdimen%
            \box0 \break}%
            \endgroup\let\graphfile=\blank}%
\def\figstop/#1#2/#3/#4#5{%
            \@midfalse\p@gefalse\begingroup%
            \mmfigcap=#3 \global\figcapht=\baselineskip%
            \global\multiply\figcapht by\mmfigcap%
               \setbox0=\hbox to\hsize{%
                \vbox to\figcapht{\graphcontent\figcapbox/L/#1/{#4}}%
                \hfill \vbox to\figcapht{\vss\figcapbox/L/#2/{#5}}}%
            \insert\topins{\penalty100 \floatingpenalty=0%
            \splittopskip=0pt \splitmaxdepth=\maxdimen%
            \box0 \break} \endgroup\let\graphfile=\blank}%
\def\figsfoot/#1#2/#3/#4#5{%
            \@midfalse\p@gefalse\begingroup%
            \mmfigcap=#3 \global\figcapht=\baselineskip%
            \global\multiply\figcapht by\mmfigcap%
               \setbox0=\hbox to\hsize{%
                \vbox to\figcapht{\graphcontent\figcapbox/L/#1/{#4}}%
                \hfill \vbox to\figcapht{\vss\figcapbox/L/#2/{#5}}}%
            \global\footsymbolcount=-1%
            \insert\footins{\penalty100 \floatingpenalty=0%
            \splittopskip=0pt \splitmaxdepth=\maxdimen%
            \box0 \break} \endgroup\let\graphfile=\blank}%
\def\figpage/#1/#2{%
             \@midfalse\p@getrue\begingroup%
             \setbox0=\vbox{\graphcontent\figcapbox/C/#1/{#2}}%
             \insert\topins{\penalty100 \floatingpenalty=0%
             \splittopskip=0pt \splitmaxdepth=\maxdimen \dimen@=\dp0%
             \vbox to\vsize{\vfill \unvbox0 \kern-\dimen@}}\endgroup%
             \let\graphfile=\blank%
             }
\def\figspage/#1#2/#3#4{%
             \@midfalse\p@getrue\begingroup%
             \setbox0=\hbox{%
               \vbox{\graphcontent\figcapbox/L/#1/{#3}}%
               \hskip 5mm%
               \vbox{\vss\figcapbox/L/#2/{#4}}}%
             \insert\topins{\penalty100 \floatingpenalty=0%
             \splittopskip=0pt \splitmaxdepth=\maxdimen \dimen@=\dp0%
             \vbox to\vsize{\vfill \box0 \kern-\dimen@}}\endgroup%
             \let\graphfile=\blank%
             }
\def\figleft/#1/#2/#3{\figcolumn/L/#1/#2/{#3}}
\def\figright/#1/#2/#3{\figcolumn/R/#1/#2/{#3}}
\def\figcenter/#1/#2/#3{%
              \@midtrue\begingroup
              \mmfigcap=#2 \global\figcapht=\baselineskip%
              \global\multiply\figcapht by\mmfigcap%
              \setbox0=\vbox
                to\figcapht{\graphcontent\figcapbox/C/#1/{#3}}
              \dimen@=\ht0
              \advance\dimen@ by\pagetotal
              \ifdim\dimen@>\pagegoal
                   \endgraf \box0 \endgraf
              \else 
                   \endgraf \box0 \endgraf\fi
              \endgroup\let\graphfile=\blank
              }
\def\figscenter/#1#2/#3/#4#5{%
               \@midtrue\begingroup
               \mmfigcap=#3 \global\figcapht=\normalbaselineskip%
               \global\multiply\figcapht by\mmfigcap%
               \global\vbadness=10000
               \setbox0=\hbox{%
                 \vbox
                   to\figcapht{\graphcontent\figcapbox/L/#1/{#4}}
                 \vbox
                  to\figcapht{\vss\figcapbox/L/#2/{#5}}}
               \if@mid \dimen@=\ht0 \advance\dimen@ by\dp0
               \advance\dimen@ by12pt \advance\dimen@ by\pagetotal
               \ifdim\dimen@>\pagegoal\@midfalse\p@gefalse\fi\fi
               \endgraf \box0 \smallbreak \endgraf \endgroup
               \let\graphfile=\blank
               }
\catcode`@=12 
%
\catcode `\@=11
\let\rel@x=\relax
\let\n@expand=\relax
\def\pr@tect{\let\n@expand=\noexpand}
\let\protect=\pr@tect
\let\gl@bal=\global
\newtoks\t@a \newtoks\t@b \newtoks\next@a
\newcount\num@i \newcount\num@j \newcount\num@k
\newcount\num@l \newcount\num@m \newcount\num@n
\long\def\l@append#1\to#2{\t@a={\\{#1}}\t@b=\expandafter{#2}%
                         \edef#2{\the\t@a\the\t@b}}
\long\def\r@append#1\to#2{\t@a={\\{#1}}\t@b=\expandafter{#2}%
                         \edef#2{\the\t@b\the\t@a}}
\def\l@op#1\to#2{\expandafter\l@opoff#1\l@opoff#1#2}
\long\def\l@opoff\\#1#2\l@opoff#3#4{\def#4{#1}\def#3{#2}}
%
%
\newif\ifnum@loop \newif\ifnum@first \newif\ifnum@last
\def\sort@@#1{\num@firsttrue\num@lasttrue\sort@t#1}
\def\sort@t#1{\pop@@#1\to\num@i\rel@x
            \ifnum\num@i=0 \num@lastfalse\let\next@a\rel@x%
            \else\num@looptrue%
                 \loop\pop@@#1\to\num@j\rel@x
                    \ifnum\num@j=0 \num@loopfalse%
                    \else\ifnum\num@i>\num@j%
                         \num@k=\num@j\num@j=\num@i\num@i=\num@k%
                         \fi%
                    \fi%
                    \push@\num@j\to#1%
                  \ifnum@loop\repeat%
                  \let\next@a\sort@t%
            \fi%
            \print@num%
            \next@a#1}
\def\print@num{%
              \ifnum@first%
                 \num@firstfalse\num@n=\num@i\number\num@i%
              \else%
                 \num@m=\num@i\advance\num@m by-\num@l%
                 \ifcase\num@m\message{%
                   *** WARNING *** Reference number %
                   [\the\num@i] appears twice or more!}%
                 \or\rel@x%
                 \else\num@m=\num@l\advance\num@m by-\num@n%
                    \ifcase\num@m\rel@x%
                    \or,\number\num@l%
                    \else-\number\num@l%
                    \fi%
                    \ifnum@last\num@n=\num@i,\number\num@i\fi%
                 \fi%
              \fi%
              \num@l=\num@i%
              }
\def\pop@@#1\to#2{\l@op#1\to\z@@#2=\z@@}
\def\push@#1\to#2{\edef\z@@{\the#1}\expandafter\r@append\z@@\to#2}


\def\append@cs#1=#2#3{\xdef#1{\csname%
                    \expandafter\g@bble\string#2#3\endcsname}}
\def\g@bble#1{}

\def\if@first@use#1{\expandafter\ifx\csname\expandafter%
                              \g@bble\string#1text\endcsname\relax}
\def\keep@ref#1#2{\def#1{0}\append@cs\y@@=#1{text}\expandafter\edef\y@@{#2}}
\def\keepref#1#2{\if@first@use#1\keep@ref#1{#2}%
                 \else\message{%
                    \string#1 is redefined by \string\keepref! %
                    The result will be .... what can I say!!}%
                 \fi}

\def\Null{0}
\def\get@ref#1#2{\def#2{\string#1text}}
\def\findref@f#1{%
                \ifx#1\Null \get@ref#1\text@cc\R@F#1{{\text@cc}}%
                \else\rel@x\fi}
\def\find@rs#1{\ifx#1\endrefs \let\next=\rel@x%
               \else\findref@f#1\r@append#1\to\void@@%
                    \let\next=\find@rs \fi \next}
\let\endrefs=\rel@x

%
\def\findref#1{\findref@f#1\ref@mark{#1}}
\def\Findref#1{\findref@f#1\Ref@mark{#1}}                            
\def\sfindref#1{\findref@f#1}                                       
%
%
\def\findrefs#1\endrefs{\def\void@@{}%
                    \find@rs#1\endrefs\r@append{0}\to\void@@%
                    \ref@mark{{\sort@@\void@@}}}
\def\Findrefs#1\endrefs{\def\void@@{}
                    \find@rs#1\endrefs\r@append{0}\to\void@@%
                    \Ref@mark{{\sort@@\void@@}}}
\def\sfindrefs#1\endrefs{\def\void@@{}%
                    \find@rs#1\endrefs\r@append{0}\to\void@@}

\newcount\referencecount     \referencecount=0
\newif\ifreferenceopen       \newwrite\referencewrite
\newdimen\refindent          \refindent=30pt
\def\REF@NUM#1{\gl@bal\advance\referencecount by 1%
    \xdef#1{\the\referencecount}}
\def\R@F#1{\REF@NUM #1\R@F@WRITE}
\def\r@fitem#1{\par \hangafter=0 \hangindent=\refindent \Textindent{#1}}
%
%
\def\refmark#1{\attach{\scriptstyle #1) }}    
\def\Refmark#1{#1)}                                             
\def\PLrefmark#1{[#1]}                        
\def\NPrefmark#1{$^{#1}$)}                    
%
%
\def\ref@mark#1#2{\if#2,\rlap#2\refmark#1%
                  \else\if#2.\rlap#2\refmark#1%
                   \else\refmark{{#1}}#2\fi\fi}
\def\Ref@mark#1#2{\if#2,\Refmark#1,
                  \else\if#2.\Refmark#1.%
                   \else\thinspace\Refmark{{#1}} #2\fi\fi}
\def\PLref@mark#1#2{\if#2,\PLrefmark#1,%
                  \else\if#2.\PLrefmark#1.%
                   \else\thinspace\PLrefmark{{#1}} #2\fi\fi}
\def\NPref@mark#1#2{\if#2,\NPrefmark#1,%
                  \else\if#2.\NPrefmark#1.%
                   \else\thinspace\NPrefmark{{#1}} #2\fi\fi}
%
%
\def\refitem#1{\r@fitem{#1.}}             
\def\NPrefitem#1{\r@fitem{#1)}}           
\def\PLrefitem#1{\r@fitem{[#1]}}          
%
\def\PTPrefs{\let\refitem=\NPrefitem}
\def\NPrefs{\let\refmark=\NPrefmark%
            \let\refitem=\NPrefitem%
            \let\ref@mark=\NPref@mark}
\def\PLrefs{\let\refmark=\PLrefmark%
            \let\Refmark=\PLrefmark%
            \let\refitem=\PLrefitem%
            \let\ref@mark=\PLref@mark}
%
%
%
\def\R@F@WRITE#1{\ifreferenceopen\else\gl@bal\referenceopentrue%
     \immediate\openout\referencewrite=\jobname.ref%
     \toks@={\begingroup \refoutspecials}%
     \immediate\write\referencewrite{\the\toks@}\fi%
     \immediate\write\referencewrite{\noexpand\refitem%
                                    {\the\referencecount}}%
    \immediate\write\referencewrite#1}

\def\outrefs{\par\penalty-400\vskip\chapterskip
   \ifreferenceopen \toks0={\par\endgroup}%
   \immediate\write\referencewrite{\the\toks0}%
   \immediate\closeout\referencewrite%
   \referenceopenfalse \fi
   \centerline{\bf References}\vskip\headskip      
   \input \jobname.ref
   }
\def\refoutspecials{\sfcode`\.=1000 \interlinepenalty=1000
         \rightskip=\z@ plus 1em minus \z@ }

%
%

\font\fourteenrm=cmr10 scaled\magstep2
\font\twelverm=cmr10 scaled\magstep1
\font\ninerm=cmr9	     \font\sixrm=cmr6
\newskip\chapterskip	     \chapterskip=\bigskipamount
\newskip\sectionskip	     \sectionskip=\medskipamount
\newskip\headskip	         \headskip=8pt plus 3pt minus 3pt
\newdimen\chapterminspace    \chapterminspace=15pc
\newdimen\sectionminspace    \sectionminspace=10pc
\newdimen\referenceminspace  \referenceminspace=25pc
\def\Textindent#1{\noindent\llap{#1\enspace}\ignorespaces}
\def\space@ver#1{\let\@sf=\empty \ifmmode #1\else \ifhmode
   \edef\@sf{\spacefactor=\the\spacefactor}\unskip${}#1$\relax\fi\fi}
\def\attach#1{\space@ver{\strut^{\mkern 2mu #1} }\@sf\ }
\def\spacecheck#1{\dimen@=\pagegoal\advance\dimen@ by -\pagetotal
   \ifdim\dimen@<#1 \ifdim\dimen@>0pt \vfil\break \fi\fi}
\catcode `\@=12
%
%
\Twelvepoint
\parindent=20 truept
\baselineskip 22 truept
\hsize=16.5 true cm
\vsize=23.2 true cm

\def\frac#1#2{ { #1 \over #2} }

\def\ket#1{| #1 \rangle}

\def\braketb#1#2{\langle #1 | #2 \rangle}
\def\braketc#1#2#3{\langle #1 | #2 | #3 \rangle}
\def\braketr#1#2#3{\langle #1 || #2 || #3 \rangle}

\def\dket#1{| #1 \rangle\!\rangle}

\def\dbraketb#1#2{\langle\!\langle #1 | #2 \rangle\!\rangle}
\def\dbraketc#1#2#3{\langle\!\langle #1 | #2 | #3 \rangle\!\rangle}

\def\rket#1{| #1 )}

\def\rbraketc#1#2#3{( #1 | #2 | #3 )}

\def\Tr#1#2{ \,t_{#1}\bigl[ #2 \bigr] }

\def\gtsim{\mathrel{\hbox{\raise0.2ex
     \hbox{$>$}\kern-0.75em\raise-0.9ex\hbox{$\sim$}}}}
\def\ltsim{\mathrel{\hbox{\raise0.2ex
     \hbox{$<$}\kern-0.75em\raise-0.9ex\hbox{$\sim$}}}}
\def\gtlt{\mathrel{\hbox{\raise0.7ex
     \hbox{$>$}\kern-0.75em\raise-0.3ex\hbox{$<$}}}}
\def\ltgt{\mathrel{\hbox{\raise0.7ex
     \hbox{$<$}\kern-0.75em\raise-0.3ex\hbox{$>$}}}}
\def\gele{\mathrel{\vbox{\hbox{\raise1.0ex
     \hbox{$\ge$}\kern-0.75em\raise-0.7ex\hbox{$<$}}}}}
\def\lege{\mathrel{\vbox{\hbox{\raise1.0ex
     \hbox{$\le$}\kern-0.75em\raise-0.7ex\hbox{$>$}}}}}

\def\dI{ {\mit\Delta I} }
\def\Ii{ I_{\rm i} }
\def\If{ I_{\rm f} }

\def\Mi{ M_{\rm i} }
\def\Mf{ M_{\rm f} }
\def\dK{ {\mit\Delta K} }
\def\Ki{ K_{\rm i} }
\def\Kf{ K_{\rm f} }
\def\cJ{ {\cal J} }
\def\cD{ {\cal D} }
\def\cM{ {\cal M} }
\def\cQ{ {\cal Q} }
\def\cT{ {\cal T} }
\def\tQ{ {\widetilde Q} }
\def\hm{ {\widehat m} }

\def\cR{ {\cal R} }
\def\cRconj{ {\cal R}{\rm -conj.} }
\def\in{ {\rm in} }

\def\unsym{ {\rm unsym} }

\def\i{ {\rm i} }
\def\f{ {\rm f} }
\def\rot{ {\rm rot} }
\def\osc{ {\rm osc} }

\def\RPA{ {\rm RPA} }

\def\A{ {\rm A} }
\def\B{ {\rm B} }
\def\a{ {\rm a} }
\def\b{ {\rm b} }
\def\c{ {\rm c} }
\def\d{ {\rm d} }
\def\g{ {\rm g} }

\def\N{ {\rm N} }
\def\n{ {\rm n} }
\def\p{ {\rm p} }

\font\smcapr=cmr8
\font\smcapi=cmti8
\def\gK{ g_{\kern-0.10em\raise-0.25ex\hbox{\smcapi K}} }
\def\gR{ g_{\kern-0.05em\raise-0.25ex\hbox{\smcapr R}} }


\keepref\Edm{ A.~R.~Edmonds, {\it Angular Momentum in Quantum Mechanics},
            (Princeton University Press, 1960).}

\keepref\BMa{ \AA.~Bohr and B.~R.~Mottelson, {\it Nuclear Structure},
             Vol. I, (Benjamin, New York, 1969), Chap. 3-3b.}

\keepref\BMb{ \AA.~Bohr and B.~R.~Mottelson, {\it Nuclear Structure},
             Vol. II, (Benjamin, New York, 1975), Chap. 4.}

\keepref\Marb{ E.~R.~Marshalek,
        Nucl. Phys. {\bf A266} (1976), 317; {\bf A275} (1977), 416.}

\keepref\SMa{ Y.~R.~Shimizu and K.~Matsuyanagi,
        Prog. Theor. Phys. {\bf 70} (1983), 144; {\bf 72} (1984), 799.}

\keepref\EMR{ J.~L.~Egido, H.~J.~Mang and P.~Ring,
        Nucl. Phys. {\bf A339} (1980), 390.}

\keepref\SMw{ Y.~R.~Shimizu and M.~Matsuzaki,
        Nucl. Phys. {\bf A588} (1995), 559.}

\keepref\MSM{ M.~Matsuzaki, Y.~R.~Shimizu and K.~Matsuyanagi,
        Prog. Theor. Phys. {\bf 79} (1988), 836.}

\keepref\NMM{ T.~Nakatsukasa, S.~Mizutori and K.~Matsuyanagi,
        Prog. Theor. Phys. {\bf 89 } (1993), 847.}

\keepref\RER{ L.~M.~Robledo, J.~L.~Egido and P.~Ring,
        Nucl. Phys. {\bf A449} (1986), 201.}

\keepref\MiSM{ S.~Mizutori, Y.~R.~Shimizu and K.~Matsuyanagi,
        Prog. Theor. Phys. {\bf 83 } (1990), 666;
         {\bf 85} (1991), 559; {\bf 86} (1991), 131.}

\keepref\NMMS{ T.~Nakatsukasa, K.~Matsuyanagi, S.~Mizutori and Y.~R.~Shimizu,
      Phys. Rev. {\bf C53} (1996), 2213.}

\keepref\Fra{ S.~Frauendorf, Nucl. Phys. {\bf A557} (1993), 259c.}

\keepref\BesT{ D.~R.~Bes and J.~Kurchan,
     {\it The Treatment of Collective Coordinates in Many-Body Systems},
     World Scientific Lecture Notes in Physics, Vol.34,
     (World Scientific, Singapore, 1990).}

\keepref\BRnil{ T.~Bengtsson and I.~Ragnarsson,
    Nucl. Phys. {\bf A436} (1985), 14.}

\keepref\NV{K.~Neeg\aa rd and P.~Vogel,
    Nucl. Phys. {\bf A145} (1970), 33;
    {\bf A149} (1970), 209; {\bf A149} (1970), 217.}

\keepref\Ucoul{ D.~Ward, H.~R.~Andrews, G.~C.~Ball, A.~Galindo-Uribarri,
  V.~P.~Janzen, T.~Nakatsukasa, D.~C.~Radford, T.~E.~Drake, J.~DeGraaf,
  S.~Pilotte and Y.~R.~Shimizu,
  Nucl. Phys. {\bf A600} (1996), 88.}

\keepref\DefDT{ K.~E.~G.~L\"obner, M.~Vettter, V.~H\"oing,
    Nucl. Data Tables, {\bf A7} (1970), 495.}

\keepref\GapDT{ G.~Audi and A.~H.~Wapstra,
    Nucl. Phys. {\bf A565} (1993), 1.}

\keepref\GdDyDT{ F.~K.~McGowan and W.~T.~Milner,
    Phys. Rev. {\bf C23} (1981), 1926.}

\keepref\ErDT{ F.~K.~McGowan, W.~T.~Milner, R.~L.~Robinson,
  P.~H.~Stelson, and Z.~W.~Grabowski,
    Nucl. Phys. {\bf A297} (1978), 51.}

\keepref\VibDT{ P.~C.~Sood, D.~M.~Headly and R.~K.~Sheline,
    At. Data Nucl. Data Tables, {\bf 47} (1991), 89.}

\keepref\NDSHe{ R.~G.~Helmer, Nucl. Data Sheet {\bf 64} (1991) 79.}

\keepref\NDSRe{ C.~W.~Reich, Nucl. Data Sheet {\bf 68} (1993) 405.}


\centerline{ \bf A New Microscopic Approach to
                 the Rotational Intensity Relations}
\centerline{ \bf --- Application of the High-Spin Cranking Formalism ---}

\vskip 10pt
\centerline{ Yoshifumi R. Shimizu and Takashi Nakatsukasa$^\dagger$ }
\vskip 8pt
\centerline{Department of Physics, Kyushu University, Fukuoka 812, Japan }
\centerline{$^\dagger$ AECL, Chalk River Laboratories,
                       Chalk River, Ontario K0J 1J0, Canada}

\baselineskip 18 truept
\bigbreak
\vskip 2 mm
\leftline{ \bf Abstract }
\vskip 2mm

  The intensity relations for electromagnetic transition rates
in the rotational coupling scheme
have been a basic tool to understand the properties of
nuclear collective rotations.
In particular the correction terms to the leading-order relation
give the response of the intrinsic motion to the Coriolis and
centrifugal forces.
We propose a simple and systematic method to calculate
the matrix elements of intrinsic operators entering the generalized
intensity relations
that uses the microscopic cranking formalism.
Examples are given to show the usefulness of the method.

\baselineskip 22 truept
\bigbreak
\vskip 2 mm
\leftline{ \bf \S1 Introduction }
\vskip 2 mm

    The cranking model is a powerful tool to study rapidly rotating nuclei:
nonlinear effects such as the alignment of quasiparticles can be
selfconsistently described from a microscopic view point.
An apparent drawback, however, is the semiclassical treatment
of the nuclear rotational motion
which assumes rotation around one
of the principal axis (usually that perpendicular to the symmetry axis).
This drawback is not necessarily a serious problem at very
high spin, but it can be a severe limitation at low spin
or in cases where the $K$-quantum number is
comparable to the total spin.

    While the assumption of rotation around a principal axis
can be lifted within the semiclassical approximation by considering
tilted-axis rotation\findref\Fra, the precise angular-momentum
algebra (i.e. the effects of Clebsch-Gordan coefficients)
has been only taken into account by e.g. angular-momentum projection.
On the other hand, in the unified model\findref\BMb,
the rotor part of the wave functions exactly takes care of
the  angular-momentum algebra.
Despite recent efforts (e.g. Ref.\Findref\BesT )
a microscopic foundation for the model remains elusive,
nevertheless,
it has been a basic and useful tool to analyze not only
energy spectra but also electromagnetic transition rates.
The problem of the model is that there has been no systematic way
to calculate the intrinsic matrix elements
from the microscopic view point.

    In this paper we propose a systematic and yet simple method
to calculate the intrinsic matrix elements
entering into the generalized
intensity relations (GIR)\findref\BMb.  This is done by starting
from the semi-microscopic cranking formalism of Ref.\Findref\Marb
and by taking a heuristic ``quantization'' procedure.
It should be noted that the method of GIR
is essentially perturbative
with respect to angular momentum\findref\BMb:
it is therefore applicable at low spin,
and in this respect, the method is complementary to
the usual cranking model which is useful at high spin.

    The paper is organized as follows:
In \S2 the basic formalism is presented.  Explicit formulae
for the $M1$ and $E\lambda$ intraband and interband
transitions, obtained as a result of application
of the basic formalism, are presented in \S3.
Some examples of numerical calculations
for the quadrupole- and octupole-vibrational bands
are discussed in \S4.
Section \S5 is devoted to the concluding remarks.

\bigbreak
\vskip 5 mm
\leftline{ \bf \S2 Basic Formalism}
\vskip 2 mm

    In this paper we assume that nuclei are axially symmetric
in their ground states
and that signature symmetry is satisfied.\footnote{*)}{
  This does not exclude the possibility of the rotationally induced
  triaxial deformation at higher spin.
  We only assume that the deformation is
  axially symmetric at rotational frequency $\omega_\rot=0$
  in the cranking terminology.}

    The GIR is obtained
by taking the matrix element of the transition operator
in the laboratory frame with respect to the unified model
wave functions, Eq.(4-19) in Ref.\Findref\BMb,
$$
   \Psi_{KIM}(\Theta,q) = \sqrt{\frac{2I+1}{(1+\delta_{K0})16\pi^2}} \,
     \Bigl( \cD^I_{MK}(\Theta) \Phi_K(q) \, + \cRconj \Bigr) , \quad
   (K \geq 0) .  \eqno(2.1)
$$
In this description, the coupling between rotational and intrinsic motion is
not seen explicitly, but it appears in the angular-momentum dependence of
the intrinsic operators.
The leading-order relation and the higher-order corrections are derived
by expanding the intrinsic operator with respect to the (intrinsic)
angular-momentum operators.
For example, the operator for the $K$-allowed transitions
($\dK \leq \lambda$) is expressed as
$$
\eqalignno{
    \cM(\lambda \mu) &= \sum_{\nu} \cM(\lambda \nu ; I_{\pm} )
              \cD^\lambda_{\mu \nu}(\Theta) \cr
      &= \sum_{\nu} \hm^{(0)}_{\lambda \nu} \, \cD^\lambda_{\mu \nu}
       + \sum_{\nu} \hm^{(+1)}_{\lambda \nu} \,
           \frac{1}{2} \bigl\{ I_+,\cD^\lambda_{\mu \nu} \bigr\}
       + \sum_{\nu} \hm^{(-1)}_{\lambda \nu} \,
           \frac{1}{2} \bigl\{ I_-,\cD^\lambda_{\mu \nu} \bigr\}
       + \,\, \cdots,   &(2.2) \cr}
$$
where the $\hm$'s are intrinsic operators which do not depend on the
rotational variables $(\Theta, I_{\pm})$.\footnote{*}{
  The intrinsic operator $\hm^{(n)}_{\lambda\nu}$ corresponds
  to the moment $m_{\nu-n,\nu}$ of Ref.\Findref\BMb.}
The transitions with definite $\dK \equiv \Kf - \Ki$
are obtained within the lowest-order corrections
by the operator,
$$
\eqalignno{
    \cM(\lambda \mu)_\dK
      =& \hm^{(0)}_{\lambda \dK} \, \cD^\lambda_{\mu \dK}
       + \hm^{(+1)}_{\lambda \dK+1} \,
           \frac{1}{2} \bigl\{ I_+,\cD^\lambda_{\mu, \dK+1} \bigr\}
       + \hm^{(-1)}_{\lambda \dK-1} \,
           \frac{1}{2} \bigl\{ I_-,\cD^\lambda_{\mu, \dK-1} \bigr\} \cr
      & \quad\quad\quad\quad
       + (1-\delta_{\dK, 0}) \,\cRconj , &(2.3\a) \cr}
$$
where for simplicity we omit the so-called
signature-dependent (decoupling) terms\findref\BMb
in the following.  Such terms can be easily evaluated if necessary.
Then the reduced transition amplitude is written as
$$
\eqalignno{
  \cT(\lambda:\i \to \f)
    & \equiv \braketr{\Kf\If}{\cM(\lambda)}{\Ki\Ii} /\sqrt{2\Ii+1} \cr
    & = C_{\i\f}\,\Bigl[ \rbraketc{\f}{\hm^{(0)}_{\lambda \dK}}{\i} \,
     \braketb{\Ii \Ki \lambda \dK}{\If \Kf} + ...\Bigr] ,  &(2.3\b) \cr}
$$
where the quantity $C_{\i\f}$ is defined as
$$
     C_{\i\f} \equiv \left\{
   \eqalign{
     1 \,\,     & \quad {\rm if} \,\,\,
         \Ki=\Kf=0   \,\,{\rm or}\,\, \Ki \ne 0,\,\Kf \ne 0,   \cr
    \sqrt{2}    & \quad {\rm if} \,\,\,
         \Ki=0,\,\Kf \ne 0 \,\,{\rm or}\,\, \Ki \ne 0,\,\Kf=0, \cr
               } \right.          \eqno(2.4)
$$
The explicit form of the lowest-order-correction terms
is not shown in Eq.(2.3b) but will be discussed later.
Thus the GIR is completely determined by the intrinsic
matrix elements, $\rbraketc{\f}{\hm^{(0)}_{\lambda \dK}}{\i} $,
$\rbraketc{\f}{\hm^{(\pm 1)}_{\lambda \dK\pm 1}}{\i} $, etc..
Needless to say, the operator
$\hm^{(\pm 1)}_{\lambda \dK\pm 1}$
with $|\dK \pm 1| > \lambda$
is understood to be vanishing, and furthermore,
the initial and final states,
$\rket{\i} $ and $\rket{\f} $,
must have good $K$-quantum numbers ($K\geq 0$).
Here and hereafter we use round brackets to denote the states and
the intrinsic matrix
elements in the unified model.

    By contrast, the semi-microscopic cranking formalism of
Ref.\Findref\Marb leads to a simpler form for
the reduced transition amplitude, Eq.(2.3b),
if it is combined with the $1/I$-expansion technique
(valid in the high-spin limit),
$$
   \cT(\lambda:\i \to \f) \approx
   \braketc{\f}{\tQ_{\lambda \mu=\dI}}{\i} ,
               \eqno(2.5)
$$
where $\dI \equiv \If - \Ii$ and
$\tQ_{\lambda \mu}$ is the transition operator
defined with respect to the rotation (cranking) axis
and the matrix element $\braketc{\f}{\tQ_{\lambda \mu}}{\i} $
is microscopically calculated by the usual cranking prescription
at finite rotational frequency $\omega_\rot$
(we use $\hbar=1$ unit in this paper).
The initial and final states in the cranking model,
$\ket{\i} $ and $\ket{\f} $, are prepared in
such a way as to guarantee a good signature-quantum number,
which is consistent with using the good-signature operator
$\tQ_{\lambda \mu=\dI}$.

The essential idea is to consider the limit,
$\omega_\rot \to 0$, which is opposite
to that normally used in the cranking formalism
and to relate the quantities appearing in the cranking calculations
to the intrinsic matrix elements in the unified model in Eq.(2.3).
At first sight this may seem impossible,
but we will show that it can be done for arbitrary types of transitions
due to a remarkable correspondence between (2.3) and (2.5)
(it has already been done for the $E2$ transition from the
$\gamma$-vibrational band to
the ground-state band in Ref.\Findref\SMw ).

     The first step in obtaining the correspondence is to correct
the difference of the basis states used in the two models:
$K$ is a good quantum number in the unified model,
while signature is good in the cranking prescription.
The states with $K \ne 0$ are degenerate at $\omega_\rot=0$,
and roughly speaking,
a pair of degenerate signature-conjugate states, say,
$\ket{\i} $ and $\ket{\bar \i} $,
form a strongly-coupled $\dI=1$ rotational band corresponding
to $\rket{\i} $.  Thus, the starting point, Eq.(2.5) should be modified:
$$
   \cT(\lambda:\i \to \f) \approx
   C_{\i\f}\,\dbraketc{\f}{\tQ_{\lambda \mu=\dI}}{\i} ,
               \eqno(2.6)
$$
where the states $\dket{\i(\f)} $ are constructed as
a linear combination of $\ket{\i(\f)} $
and it's signature-conjugate states $\ket{\bar \i (\bar \f)} $
so as to make
$K$ a good quantum number in the limit $\omega_\rot =0$.
Since $\ket{\i} $ and $\ket{\bar \i} $ are independent,
the relative phase between them is free within the cranking formalism.
We fix the phase in this paper by,
$$
  \Ki = \braketc{\bar \i}{J_z}{\i} = \braketc{\i}{J_z}{\bar \i}
      \geq 0 , \quad\quad
      {\rm at}\quad \omega_\rot=0 .  \eqno(2.7\a)
$$
Then the $\dket{\i}$ is explicitly defined by
$$
  \left\{
  \eqalign{
  \dket{\i} = \ket{\i}
       \hskip3.2cm & {\rm for}\,\, \Ki=0 , \cr
  \dket{\i} =
         \frac{1}{\sqrt{2}} \bigl( \ket{\i} + \ket{\bar \i} \bigr)
          \quad\quad\quad & {\rm for}\,\, \Ki > 0 , \cr
  \dket{-\i} =
         \frac{1}{\sqrt{2}} \bigl( \ket{\i} - \ket{\bar \i} \bigr)
           \quad\quad\quad & {\rm for}\,\, \Ki < 0 , \cr
               } \right.     \eqno(2.7\b)
$$
which has definite $\Ki$ values in the $\omega_\rot=0$ limit
(the same for $\dket{\f} $).
We call a set of states defined in this way
a ``$K$-good'' representation, though they do not have good $K$
quantum numbers at finite $\omega_\rot$.
In the following,
we use $\dket{\i}$ ($\dket{\f}$) for states with non-negative $K$-values
and $\dket{-\i}$ ($\dket{-\f}$) for those with negative $K$.
The coefficient $C_{\i\f}$ in Eq.(2.6),
which is defined by Eq.(2.4)
through the $\Ki$ and $\Kf$ values at $\omega_\rot=0$,
is introduced in order to guarantee that
Eq.(2.6) recovers the original Eq.(2.5) in the high-spin limit.

     The next step is to decompose
the multipole transition operators $\tQ_{\lambda \mu}$ in Eq.(2.6), which
are defined with respect to the rotation axis ($x$-axis),
into those defined with respect to the symmetry axis ($z$-axis),
$Q_{\lambda \nu}$:
$$
  \tQ_{\lambda \mu} =  \sum_{\nu}
      \cD^\lambda_{\mu \nu} \Bigl(-\frac{\pi}{2}, -\frac{\pi}{2}, 0 \Bigr)
           \,  Q_{\lambda \nu}
      =   i^{-\mu} \sum_{\nu} d^\lambda_{\mu\nu} \Bigl(-\frac{\pi}{2} \Bigr)
           \,  Q_{\lambda \nu},       \eqno(2.8)
$$
where we followed Ref.\Findref\BMb for definitions of
the $\cD$ and $d$ functions.
The original GIR was obtained by expanding the matrix elements
with respect to the angular momentum.  Correspondingly,
the matrix elements in the cranking prescription are expanded
with respect to the rotational frequency:
$$
     \dbraketc{\f}{Q_{\lambda \nu}}{\i} =\left\{
       \eqalign{
           O(1) \quad       & \quad {\rm for} \quad \nu=\dK , \cr
           O(\omega_\rot)   & \quad {\rm for} \quad \nu=\dK \pm 1 , \cr
           O(\omega^2_\rot) & \quad {\rm otherwise} , \cr
               } \right.          \eqno(2.9)
$$
because the cranking term mixes the states with $\dK=\pm 1$.
Using Eqs.(2.8) and (2.9),
we obtain
within the lowest-order corrections with respect to $\omega_\rot$,
$$
\eqalignno{
  \dbraketc{\f}{\tQ_{\lambda \mu=\dI}}{\i}
      &= i^{-\dI}
     \Biggl(
  \Bigl[ \dbraketc{\f}{Q_{\lambda \dK}}{\i} \Bigr]_0
         d^\lambda_{\dI \dK} \Bigl(-\frac{\pi}{2}\Bigr) \cr
     &\quad\quad\quad +
  \Bigl[ \frac{d\dbraketc{\f}{Q_{\lambda \dK+1}}{\i}}{d\omega_\rot} \Bigr]_0
   \omega_\rot \, d^\lambda_{\dI \dK+1} \Bigl(-\frac{\pi}{2}\Bigr) \cr
     &\quad\quad\quad +
  \Bigl[ \frac{d\dbraketc{\f}{Q_{\lambda \dK-1}}{\i}}{d\omega_\rot} \Bigr]_0
   \omega_\rot \, d^\lambda_{\dI \dK-1} \Bigl(-\frac{\pi}{2}\Bigr)
     \Biggr) . &(2.10) \cr}
$$
Here $[*]_0$ means that the expression is evaluated
by taking the limit $\omega_\rot \to 0$.

The key to obtain the correspondence is the $d$ function,
which has an asymptotic form\findref\Edm,
$$
     d^\lambda_{\dI\dK}(-\theta) \approx
          \braketb{\Ii \Ki \lambda \dK}{\If \Kf} , \eqno(2.11\a)
$$
with
$$
      \dI \equiv \If-\Ii , \,\,  \dK \equiv \Kf-\Ki , \quad
        \cos{\theta} \approx \Ki/\Ii \approx \Kf/\If ,  \eqno(2.11\b)
$$
which is valid for integer $\lambda$ and for
$\dI,\dK \ll \Ii,\If$.
The cranking prescription, Eq.(2.6) with Eq.(2.8), takes the
high-spin limit $K \ll I$, namely $\theta \approx \frac{\pi}{2}$.
Since we are considering the opposite limit ($\theta < \frac{\pi}{2}$),
it is natural to replace the $d$ function of the leading-order term
in Eq.(2.10) by the Clebsch-Gordan coefficient (2.11a).
Then the leading-order matrix element in the cranking prescription (2.6)
turns out to have the same structure as the one in the unified model,
Eq.(2.3).
An alternative interpretation of
this replacement is that
the $d$ function $d^\lambda_{\dI\dK}$ in Eq.(2.10) corresponds to
the reduced matrix element of the $\cD$ operator in Eq.(2.3)
with respect to the rotor part of unsymmetrized wave functions
in Eq.(2.1): i.e.
$$
   d^\lambda_{\dI\dK} \Bigl(-\frac{\pi}{2} \Bigr)
       \,\leftrightarrow\,\,
     \Bigl[ \frac{2\If+1}{2\Ii+1} \Bigr]^{\frac{1}{2}}\,\,
 \frac{\braketc{\Kf\If\Mf}{\cD^\lambda_{\mu \dK}}{\Ki\Ii\Mi}_{\rm unsym}}
           {\braketb{\Ii\Mf\lambda\mu}{\If\Mf}} .    \eqno(2.12\a)
$$
Note that this replacement may be regarded
as a kind of ``quantization'' incorporating
the full quantum-mechanical effects of angular-momentum algebra
present in the GIR but neglected
in the semiclassical cranking model.

     With the close relationship between the $\omega_\rot$ expansion
in the cranking calculation and the $I$ expansion in the GIR,
we may introduce a ``quantization''
for the higher-order term: i.e.
$$
   \omega_\rot \, d^\lambda_{\dI \dK \pm 1} \Bigl(-\frac{\pi}{2} \Bigr)
       \,\leftrightarrow\,\,
     \Bigl[ \frac{2\If+1}{2\Ii+1} \Bigr]^{\frac{1}{2}}\,\,
      \frac{\braketc{\Kf\If\Mf}
          {\,{\displaystyle \frac{1}{\cJ} \, \frac{1}{2} }
      \{I_{\pm}, \cD^\lambda_{\mu, \dK \pm 1}\}\,}{\Ki\Ii\Mi}_{\rm unsym}}
           {\braketb{\Ii\Mf\lambda\mu}{\If\Mf}} ,   \eqno(2.12\b)
$$
where $\cJ$ is the moment of inertia of the band under consideration
at $\omega_\rot=0$.
Noting that $\omega_\rot \approx I_x/\cJ$,
the ``quantization'' of Eq.(2.12b), i.e. the appearance of
${\displaystyle \frac{1}{\cJ} \, \frac{1}{2} }
\{I_{\pm}, \cD^\lambda_{\mu \nu} \}$
on the right hand side, is natural, though the ordering between
$I_\pm$ and $\cD^\lambda_{\mu\nu}$ is not trivial.
As in any kind of quantization procedures, the ordering of the operators
cannot be determined from classical considerations.
We have found  that the symmetrized ordering as given in Eq.(2.12b)
is the most appropriate one and it is discussed further in Appendix A.
Therefore we take this ordering throughout the present paper.
Concerning the moment of inertia, there is ambiguity because
its values are generally different for the initial and final states.
However, since the difference is small in most cases,
we may say that
$$
  \cJ \equiv \frac{1}{2} \Bigl[
    \frac{d\dbraketc{\i}{J_x}{\i}}{d\omega_\rot} +
    \frac{d\dbraketc{\f}{J_x}{\f}}{d\omega_\rot}
       \Bigr]_0  \approx \Bigl[
    \frac{d\dbraketc{\i}{J_x}{\i}}{d\omega_\rot}
       \Bigr]_0  \approx \Bigl[
    \frac{d\dbraketc{\f}{J_x}{\f}}{d\omega_\rot}
       \Bigr]_0 .            \eqno(2.13)
$$

    Using these quantization conditions, Eq.(2.12),
the correspondence between Eq.(2.3) and Eq.(2.6) with (2.10) is apparent
and the cranking procedure gives the intrinsic
matrix elements in the unified model within the lowest-order
corrections by
$$
\eqalignno{
   \rbraketc{\f}{\hm^{(0)}_{\lambda \dK}}{\i} &=
     \Bigl[ \dbraketc{\f}{Q_{\lambda \dK}}{\i} \Bigr]_0 , &(2.14\a) \cr
   \rbraketc{\f}{\hm^{(\pm 1)}_{\lambda \dK \pm 1}}{\i} &=
       \frac{1}{\cJ} \Bigl[
     \frac{d\dbraketc{\f}{Q_{\lambda \dK \pm 1}}{\i}}{d\omega_\rot}
        \Bigr]_0 .        &(2.14\b) \cr}
$$
Here and hereafter we omit the overall phase $i^{-\dI}$
which comes from Eq.(2.10).

     Although the procedure is, in principle, applicable to any order
in angular momentum, it becomes much more difficult to deduce
the general correspondence because the
number of possible orderings increases.  The only exception is
the $K$-forbidden transitions\findref\BMb.  For the transition with
the order of $K$-forbiddenness, $n = |\dK| - \lambda$ ($\ge 0$),
the operator has the form
$$
\eqalignno{
  \cM(\lambda \mu)_{\dK=\pm (n+\lambda)}
    =& \hm^{(\mp n)}_{\lambda, \pm \lambda} \,
          I^n_{\mp} \, \cD^\lambda_{\mu, \pm \lambda}
      + \hm^{(\mp (n+1))}_{\lambda, \pm (\lambda-1)} \,
   \frac{1}{2} \bigl\{ I^{n+1}_{\mp},
             \cD^\lambda_{\mu, \pm (\lambda-1)} \bigr\} \cr
    &\quad\quad + \cRconj . &(2.15) \cr}
$$
In this case, the $\omega_\rot$-expansion, Eq.(2.9),
gives the leading-order matrix elements,
$\dbraketc{\f}{Q_{\lambda, \pm \lambda}}{\i} = O(\omega^n_\rot)$,
and the first-order correction,
$\dbraketc{\f}{Q_{\lambda, \pm (\lambda - 1)}}{\i} = O(\omega^{n+1}_\rot)$.
Then the ``quantization'' with the symmetrized ordering as in Eqs.(2.12)
leads to the correspondence
$$
\eqalignno{
    \rbraketc{\f}{\hm^{(\mp n)}_{\lambda, \pm \lambda}}{\i}
      &= \frac{1}{n! \cJ^n}
       \Bigl[ \frac{d^n\dbraketc{\f}{Q_{\lambda, \pm \lambda}}{\i}}
                {d\omega_\rot^n} \Bigr]_0 ,   &(2.16\a) \cr
    \rbraketc{\f}{\hm^{(\mp (n+1))}_{\lambda, \pm (\lambda-1)}}{\i}
      &= \frac{1}{(n+1)! \cJ^{n+1}}
     \Bigl[ \frac{d^{n+1}\dbraketc{\f}{Q_{\lambda, \pm (\lambda-1)}}{\i}}
                {d\omega_\rot^{n+1}} \Bigr]_0 .  &(2.16\b) \cr}
$$

      Note that there exists a one-to-one correspondence
between the unified-model intrinsic state $\rket{\i}$ and
the cranking-model state $\dket{\i}$
because of the introduction of the $K$-good representation (2.7)
at $\omega_\rot=0$.
{}From Eqs.(2.14) and (2.16),
it may be generally written
$$
    \rbraketc{\f}{\hm^{(n)}_{\lambda \nu}}{\i}
     \equiv  \frac{1}{|n|!\cJ^{|n|}} \Bigl[
     \frac{d^{|n|}\dbraketc{\f}{Q_{\lambda \nu}}{\i}}{d\omega^{|n|}_\rot}
        \Bigr]_0 ,   \quad {\rm for}\,\,\, \dK=\nu-n . \eqno(2.17)
$$
With this simple prescription,
we are able to calculate microscopically
the intrinsic matrix elements
in the unified model by using the cranking model.

Although they are not considered explicitly in the following examples,
it is worthwhile mentioning that
signature-dependent contributions (decoupling terms) can be
evaluated in the same way as in the original treatment of the GIR;
namely the intrinsic matrix elements are
$$
    \rbraketc{\f}{\hm^{(n)}_{\lambda\nu}\cR^{-1}_i}{\i}
    =(-)^{K_\i} \,\rbraketc{\f}{\hm^{(n)}_{\lambda\nu}\cR^{-1}_x}{\i}
    =(-)^{\alpha_\i+K_\i} \frac{1}{|n|!\cJ^{|n|}} \Bigl[
     \frac{d^{|n|}\dbraketc{\f}{Q_{\lambda \nu}}{-\i}}{d\omega^{|n|}_\rot}
        \Bigr]_0 ,        \eqno(2.18)
$$
where the intrinsic $\cR$-conjugation operator\findref\BMb is
$\cR_i=\exp{(-i\pi J_y)}$, while the signature-conjugation operator
is $\cR_x=\exp{(-i\pi J_x)}$, and $\alpha_\i$ is the
signature-quantum number of the cranking state $\ket{\i}$,
i.e. $\cR_x\ket{\i}=(-1)^{-\alpha_\i}\,\ket{\i}$.

     In the unified model\findref\BMb,
it is straightforward to obtain the GIR from
the wave function (2.1) and the transition operator (2.2).
The main result of the present paper is the description of a
microscopic method to calculate the intrinsic matrix elements
in the $\omega_\rot \rightarrow 0$ limit
of the cranking model.

\bigbreak
\vskip 2 mm
\leftline{ \bf \S3 Some Applications}
\vskip 2 mm

   In the following, we take concrete examples of various
types of transitions and show explicitly the formula to calculate the GIR
in terms of the signature-good matrix elements of the cranking model.
We assume the usual phase convention\findref\BMa for the matrix elements
of electromagnetic transition operators.

     Since the eigenstates in cranking calculations
have good signature,
it is useful to introduce the signature-classified-transition operators,
$$
  Q^{(\pm)}_{\lambda K} \equiv \frac{1}{\sqrt{2(1+\delta_{K0})}}
   \bigl( Q_{\lambda K} \pm (-)^\lambda Q_{\lambda, -K} \bigr) ,
     \quad\quad  K \ge 0 ,     \eqno(3.1\a)
$$
where the sign $+(-)$ means that the operator transfers
the signature by $r= +1(-1)$, or $\alpha= 0\,(1)$.
Conversely,
$$
  Q_{\lambda, \pm K} = \frac{1}{\sqrt{2-\delta_{K0}}} \left\{
       \eqalign{
      \bigl( Q^{(+)}_{\lambda K} \pm Q^{(-)}_{\lambda K} \bigr) ,
           & \quad {\rm for} \quad \lambda={\rm even} , \cr
      \bigl( Q^{(-)}_{\lambda K} \pm Q^{(+)}_{\lambda K} \bigr) ,
           & \quad {\rm for} \quad \lambda={\rm odd} . \cr
               } \right.          \eqno(3.1\b)
$$
Only one signature component exists for the $K=0$ operator, namely
the one with $(+)$ for $\lambda$=even and $(-)$ for $\lambda$=odd.
Using Eqs.(2.7) and (3.1), one can easily convert the matrix elements
in the $K$-good states, $\dbraketc{\f}{Q_{\lambda K}}{\i}$,
to those in the signature-good states, e.g.
$\braketc{\f}{Q^{(+)}_{\lambda K}}{\i}$ and
$\braketc{\bar\f}{Q^{(-)}_{\lambda K}}{\i}$ etc.,
which are used in the cranking calculations.

\vskip 5 mm
\leftline{\it 3-1. Intraband M1 and E2 transitions}
\vskip 2 mm

    Let us first consider the simplest example
of the in-band $M1$ transition.
In this case the initial and final states are the same,
$\ket{\i} = \ket{\f} $, $\dK=0$ and $\Ki=\Kf\equiv K\neq 0$.
We denote these states
as $\ket{K} $ for simplicity.
The magnetic dipole operator $\vec \mu$ is classified as
$$
    \mu^{(-)}_{10}=\mu_z, \,\,\, \mu^{(+)}_{11}=-\mu_x,
     \,\,\, {\rm and}\,\,\, \mu^{(-)}_{11}=-i\mu_y.
$$
Then the leading-order matrix element (2.14a) is
$$
    \rbraketc{\f}{\hm^{(0)}_{10}}{\i}
     = \Bigl[ \dbraketc{K}{\mu_z}{K} \Bigr]_0
     = \gK K , \quad
      \gK \equiv \Bigl[
      \frac{\braketc{\bar K}{\mu_z}{K} }{\braketc{\bar K}{J_z}{K} }
             \Bigr]_0 ,   \eqno(3.2\a)
$$
where $\braketc{\bar K}{\mu_z}{K} = \braketc{K}{\mu_z}{\bar K}$
and the phase conventions (2.7) are used.
In the case of $M1$ transitions,
the first-order-correction term is
an important contribution of the same order
as the leading term\findref\BMb.  In fact, by using Eq.(2.14b)
the first-order matrix element is
$$
\eqalignno{
    \rbraketc{\f}{\hm^{(\pm 1)}_{1, \pm 1}}{\i}
    &= \mp \frac{1}{\sqrt{2}\cJ}
      \Bigl[ \frac{d\dbraketc{K}{\mu_x}{K}}{d\omega_\rot} \Bigr]_0
     = \mp \frac{1}{\sqrt{2}} \, \gR , \cr
   \gR &\equiv \Bigl[ \Bigl(
         \frac{d\braketc{K}{\mu_x}{K}}{d\omega_\rot}
       + \frac{d\braketc{\bar K}{\mu_x}{\bar K}}{d\omega_\rot}
          \Bigr)/\Bigl(
         \frac{d\braketc{K}{J_x}{K}}{d\omega_\rot}
       + \frac{d\braketc{\bar K}{J_x}{\bar K}}{d\omega_\rot}
          \Bigr) \Bigr]_0 , &(3.2\b) \cr}
$$
where we have used $\dbraketc{K}{i\mu_y}{K}=0$,
and $\ket{K}=\ket{\bar K}$ for $K=0$ bands.
Then, by using the identity,
$$
    \frac{1}{\sqrt{2}}
 \Bigl( -\frac{1}{2}\bigl\{I_+, \cD^1_{\mu, +1}\bigr\}
        +\frac{1}{2}\bigl\{I_-, \cD^1_{\mu, -1}\bigr\} \Bigr)
       = I^{\rm (lab)}_\mu - I_0 \cD^1_{\mu 0} ,\quad
   I^{\rm (lab)}_\mu \equiv \sum_\nu \cD^1_{\mu \nu} I_\nu ,
$$
and the fact that $I_0$ gives $K$ when operated, it is easy to see that
$$
   \cT(M1:\i \to \f)_\in
    =   \gR \sqrt{\If(\If+1)} \,\delta_{\Ii\If}
       + \braketb{\Ii K 1 0 }{\If K} \, (\gK - \gR)K .
       \eqno(3.3)
$$
Thus the well-known formula for the $M1$ matrix elements is derived,
where the $g$ factors, $\gK$ as well as $\gR$,
can be calculated from the microscopic
cranking formalism at the infinitesimal rotational frequency point.

    The second simple example is the in-band $E2$ transitions.
In this case the leading-order matrix element (2.14a) becomes
the quadrupole moment:
$$
\eqalignno{
   \rbraketc{\f}{\hm^{(0)}_{20}}{\i}
   & = \Bigl[ \dbraketc{K}{Q^{(+)}_{20}}{K} \Bigr]_0 = Q_0, \cr
    Q_0 &\equiv \Bigl[ \frac{1}{2} \Bigl(
     \braketc{K}{Q^{(+)}_{20}}{K}+  \braketc{\bar K}{Q^{(+)}_{20}}{\bar K}
       \Bigr) \Bigr]_0
      = \Bigl[ \braketc{K}{Q^{(+)}_{20}}{K} \Bigr]_0 . &(3.4\a) \cr}
$$
The lowest-order correction term, which is linear in angular momentum,
is non-vanishing if $K \ne 0$; by using Eq.(2.14b),
$$
   \rbraketc{\f}{\hm^{(\pm 1)}_{2, \pm 1}}{\i} = \pm Q'_1, \quad
      Q'_1 \equiv \frac{1}{\sqrt{2}\cJ} \Bigl[
     \frac{d\braketc{\bar K}{Q^{(-)}_{21}}{K}}{d\omega_\rot}
       \Bigr]_0\,(1-\delta_{K0}) .     \eqno(3.4\b)
$$
Here we have used $\dbraketc{K}{Q^{(+)}_{21}}{K}=0$ and
$\braketc{\bar K}{Q^{(-)}_{21}}{K}=\braketc{K}{Q^{(-)}_{21}}{\bar K}$.
Then the GIR is written as
$$
\eqalignno{
   \cT(E2:\i \to \f)_\in
    & = \braketb{\Ii K 2 0 }{\If K} \, Q_0 \cr
    &\quad+\Bigl(
       \braketb{\Ii K 2 1}{\If K+1} \sqrt{(\If-K)(\If+K+1)} \cr
    &\quad\quad
         -\braketb{\Ii K 2 -1}{\If K-1} \sqrt{(\If+K)(\If-K+1)}
              \Bigr) \, Q'_1 ,   &(3.5) \cr}
$$
where the lowest-correction term can be written
in various alternative ways:\findref\BMb
We have used the identity,
$$
    \frac{1}{2} \biggl(
       \bigl\{I_+, \cD^2_{\mu,+1} \bigr\} -
       \bigl\{I_-, \cD^2_{\mu,-1} \bigr\} \biggr) =
           I_+ \,\cD^2_{\mu,+1} - I_- \,\cD^2_{\mu,-1} .
$$

   The first and second terms in the right hand side of Eq.(3.5) come from
the matrix elements with signature $\alpha=0$ and $\alpha=1$, respectively.
It is interesting to verify the selection rule, $\dI = \alpha$ ($mod\, 2$),
which is valid in the high-spin cranking formalism.
In fact, the ratio of these terms takes the asymptotic value,
$$
 \frac{\rm \alpha=0\ term}{\rm \alpha=1\ term}
    \approx \frac{Q_0 } { I Q'_1}  \times \left\{
       \eqalign{
   \frac{1}{2\sqrt{6}} (K/I)^{-1}  & \quad {\rm for} \quad \dI = \pm 2 , \cr
  - \frac{\sqrt{6}}{2} (K/I) \quad & \quad {\rm for} \quad \dI = \pm 1 , \cr
   \frac{\sqrt{6}}{4}  (K/I)^{-1}  & \quad {\rm for} \quad \dI = 0, \cr
               } \right.
$$
in the limit of $\If \approx \Ii \approx I \gg K \gg 1$.
Thus, the correspondence between signature and angular momentum
in the cranking formalism is recovered at the high-spin limit.

  For the one-quasiparticle states, $Q'_1$ is
a single-particle matrix element, while $Q_0$ is the collective
matrix element.  Therefore the $Q'_1$ is typically two to three
orders of magnitude smaller than $Q_0$
and gives a negligible contribution in our formalism.
This is in agreement with experimental observations\findref\BMb.

\bigbreak
\vskip 2 mm
\leftline{\it 3-2. $K$-allowed transitions}
\vskip 2 mm

      It is straightforward, but tedious, to derive the GIR
for arbitrary transitions in the general case
because the number of non-zero matrix element
increases due to many possible ways of transferring
the $K$-quantum number,
and because of the signature distinction.
However, a rather simple formula can be obtained
for the general $\lambda$-pole transitions
if we neglect the signature-dependent contributions
and keep the matrix elements in the $K$-good representation
as in Eq.(2.17).
The GIR can then be written in different ways
depending on which Clebsch-Gordan coefficients are included.
A simple form of the GIR, Eq.(4-98) of Ref.\Findref\BMb,
was derived assuming $\dK =\Kf-\Ki \ge \lambda$.
In this section, we present the formula for
arbitrary $K$-allowed transitions with $|\dK| \le \lambda$.

     Since the signature-dependent terms are neglected there exist
two first-order-correction terms associated with the rotor part
operator,
$\frac{1}{2} \bigl\{ I_+,  \cD^\lambda_{\mu,\dK+1} \bigr\}$ and
$\frac{1}{2} \bigl\{ I_-,  \cD^\lambda_{\mu,\dK-1} \bigr\}$
as is seen in Eq.(2.3a).  Let us call them simply $I_+$ and $I_-$ terms,
respectively.  It is always possible to eliminate one of them
by using a general identity between the angular momenta
and the $\cD$ function\findref\BMb,
$$
\eqalignno{
    [(\lambda - \nu)(\lambda + \nu +1)]^{\frac{1}{2}} \,
        \frac{1}{2} \bigl\{ I_+,\cD^\lambda_{\mu, \nu +1} \bigr\}
   &+ [(\lambda + \nu)(\lambda - \nu +1)]^{\frac{1}{2}} \,
        \frac{1}{2} \bigl\{ I_-,\cD^\lambda_{\mu, \nu -1} \bigr\} \cr
   &= [ {\mib I}^2, \cD^\lambda_{\mu \nu} ]
      - \nu \bigl\{ I_0,\cD^\lambda_{\mu \nu} \bigr\} .  &(3.6) \cr}
$$
For $-\lambda \le \nu=\dK \le 0$
($\lambda \ge \nu=\dK \ge 0$) we can eliminate the $I_+$ ($I_-$) term
so as to obtain a simple form of GIR.
In order to further simplify the relation we apply the identity,
$$
  \frac{1}{2} \bigl\{ I_\pm,  \cD^\lambda_{\mu,\nu \pm 1} \bigr\}
    = I_\pm \cD^\lambda_{\mu,\nu \pm 1} - \frac{1}{2}
     \sqrt{(\lambda \mp \nu)(\lambda \pm \nu+1)} \,\cD^\lambda_{\mu\nu} ,
       \eqno(3.7)
$$
to the remaining $I_-$ ($I_+$) term.
Then the GIR for the $\lambda$-pole transitions is written as
$$
\eqalignno{
  &\cT(\lambda:\i \to \f)_\dK
    =\braketb{\Ii \Ki \lambda \dK }{\If \Kf} \, Q_t
     \, \bigl(\, 1 + q \, [\If (\If+1) - \Ii (\Ii+1)] \,
         \bigr) \cr
  &\quad\quad +\sqrt{(\If \pm \Kf)(\If \mp \Kf+1)} \,
     \braketb{\Ii \Ki \lambda \,(\dK \mp 1)}{\If \,(\Kf \mp 1)} \,
         Q'_t , \quad (\dK \ltgt 0) ,  &(3.8) \cr}
$$
where the upper (lower) sign is for $\dK \le 0$ ($\dK \ge 0$), and
the intrinsic parameters $Q_t$, $Q'_t$ and $q$ are calculated in
the $\omega_\rot \rightarrow 0$ limit of the cranking formalism by means
of Eq.(2.17);
$$
\eqalignno{
   Q_t =& C_{\i\f} \Biggl\{
   \Bigl[ \dbraketc{\f}{Q_{\lambda \dK}}{\i} \Bigr]_0
    - \frac{\sqrt{(\lambda \pm \dK)(\lambda \mp \dK+1)}}{2\cJ} \Bigl[
      \frac{ d\dbraketc{\f}{Q_{\lambda,\dK \mp 1}}{\i} }{d\omega_\rot}
        \Bigr]_0 \cr
     &\quad\quad\quad
    + \frac{(\lambda \pm \dK)(\lambda \mp \dK+1)-2\dK(\Ki+\Kf)}
             {\sqrt{(\lambda \mp \dK)(\lambda \pm \dK+1)}} \,
         \frac{1}{2\cJ} \Bigl[
      \frac{ d\dbraketc{\f}{Q_{\lambda,\dK \pm 1}}{\i} }{d\omega_\rot}
    \Bigr]_0 \Biggr\}, \cr
  Q'_t =& C_{\i\f} \Biggl\{ \frac{1}{\cJ} \Bigl[
      \frac{ d\dbraketc{\f}{Q_{\lambda,\dK \mp 1}}{\i} }{d\omega_\rot}
         \Bigr]_0
  - \sqrt{\frac{(\lambda \pm \dK)(\lambda \mp \dK+1)}
               {(\lambda \mp \dK)(\lambda \pm \dK+1)}}
       \,\frac{1}{\cJ} \Bigl[
      \frac{ d\dbraketc{\f}{Q_{\lambda,\dK \pm 1}}{\i} }{d\omega_\rot}
          \Bigr]_0 \Biggr\} , \cr
    q =& C_{\i\f} \,\frac{1}{\sqrt{(\lambda \mp \dK)(\lambda \pm \dK+1)}}
         \frac{1}{\cJ} \Bigl[
      \frac{ d\dbraketc{\f}{Q_{\lambda,\dK \pm 1}}{\i} }{d\omega_\rot}
    \Bigr]_0 \, / Q_t , \quad (\dK \ltgt 0) .  &(3.9) \cr}
$$
Note that $Q'_t$ vanishes when $\dK= \mp \lambda$
and the simple GIR\findref\BMb is recovered.
By using Eqs.(3.6) and (3.7)
it is easy to check that the formula for $\dK \le 0$
and $\dK \ge 0$ give identical results when $\dK=0$.
The results of \S3-1 can be
derived from the general formula above,
though their appearance is different at first sight.
In Eq.(3.8),
$Q_t$ is, roughly speaking, the leading-order term
and the $Q'_t$ and $q$ terms give
the lowest-order corrections in the sense of $I$ expansion.
It should be mentioned that the higher-order terms
in the main amplitude $Q_t$, Eq.(3.9), come from
the ``contraction'' due to the specific ordering expressed in Eq.(2.12),
which should be small compared to the main term,
but they do give important contributions.

\bigbreak
\vskip 2 mm
\leftline{\it 3-3. $K$-forbidden transitions}
\vskip 2 mm

    The simple GIR, Eq.(4-98) in Ref.\Findref\BMb,
can be obtained for the $K$-forbidden transitions.
Again, we consider both the $\dK < 0$ and $\dK > 0$ cases.
Using the order of $K$-forbiddenness, $n = |\dK| - \lambda > 0$,
and applying Eq.(2.17) to Eqs.(2.15) and (2.16), we obtain
{\vskip -1.0 truecm}
$$
\eqalignno{
   \cT(\lambda:\i \to \f)_{\dK=\mp (n+\lambda)}
    &= \sqrt{
   \frac{(\If \mp \Kf)!\,(\If \pm \Kf+n)!}
        {(\If \mp \Kf-n)!\,(\If \pm \Kf)!}} \,
    \braketb{\Ii \Ki \lambda \,(\mp \lambda)}{\If \,(\Kf \pm n)} \, Q_t \cr
     &\quad\quad\times \bigl( 1 + q \, [\If (\If+1) - \Ii (\Ii+1)] \,
          \bigr) , \quad (\dK \ltgt 0) ,   &(3.10) \cr}
$$
where
$$
\eqalignno{
   Q_t =& C_{\i\f} \Biggl\{ \frac{1}{n! \cJ^n}
      \Bigl[ \frac{d^n\dbraketc{\f}{Q_{\lambda,\mp \lambda}}{\i}}
                {d\omega_\rot^n} \Bigr]_0
       \pm \sqrt{\frac{\lambda}{2}} \, \frac{(\Ki+\Kf)}{(n+1)! \cJ^{n+1}}
     \Bigl[ \frac{d^{n+1}\dbraketc{\f}{Q_{\lambda, \mp (\lambda-1)}}{\i}}
             {d\omega_\rot^{n+1}}
          \Bigr]_0 \Biggr\} ,  \cr
     q =& C_{\i\f} \frac{1}{\sqrt{2\lambda}\,(n+1)! \cJ^{n+1}}
     \Bigl[ \frac{d^{n+1}\dbraketc{\f}{Q_{\lambda, \mp (\lambda-1)}}{\i}}
         {d\omega_\rot^{n+1}} \Bigr]_0 \,/Q_t , \quad (\dK \ltgt 0) .
           &(3.11)  \cr}
$$

   Once the initial and final states are specified, the intrinsic
parameters (3.11)
can be calculated in the cranking model.
This formula may find application to analyzing
the direct decays of the high-$K$ isomer states
to the ground-state band,
which have been systematically observed recently.
For two-quasiparticle high-$K$ isomers,
both $M1$ and $E2$ matrix elements are straightforward
to evaluate within the independent quasiparticle approximation.
Some correlations between the quasiparticles
will be necessary in order to obtain non-zero amplitudes
for three- or four-quasiparticle isomers.

\bigbreak
\vskip 2 mm
\leftline{\it 3-4. Vibrational transitions
to the ground-state band of even-even nuclei}
\vskip 2 mm

     One of the important applications of the GIR is to
transitions from collective vibrational bands to
the ground-state band.
Collective vibrational motion built on the ground band
can be consistently described
within the cranking model by employing the
random-phase approximation (RPA)
in the rotating frame\findrefs\Marb\EMR\SMa\RER\MiSM\NMMS\endrefs,
which we will call the simple RPA method hereafter.
In this subsection,
we discuss the GIR for vibrational transitions in even-even
nuclei with the simple RPA method.

     Since $\Kf=0$ for the ground-state band of even-even nuclei,
we have $\Ki=K$ and $\dK = -K \le 0$ for the $\lambda$-pole transitions
from the $\lambda'$-pole vibrational mode
with definite $K$ ($0 \le K \le \lambda'$)
in the $\omega_\rot=0$ limit.  Moreover,
in the simple RPA method,
the initial and final states are simply related by
$ \ket{\i} = X_K^{(\pm)\dagger} \ket{\f}$,
where $\ket{\f}=\ket{0}$ denote
the ground state rotational band and $X_K^{(\pm)\dagger}$
is the creation operator of the vibrational mode with the
signature $r=\pm 1$ and the $K$-quantum number $K$ at
$\omega_\rot=0$.  The RPA transition matrix elements
between the one-phonon states and the vacuum state at an arbitrary
rotational frequency are defined by
$$
    \Tr{K}{Q^{(\pm)}_{\lambda \nu}} \equiv
     \braketc{0}{[Q^{(\pm)}_{\lambda \nu},X_K^{(\pm)\dagger}]}{0}
   = \braketc{\f}{Q^{(\pm)}_{\lambda \nu}}{\i}^{\,}_\RPA ,
          \quad (K,\nu \ge 0) .  \eqno(3.12)
$$
Note that only the signature component $r=(-1)^\lambda$
($(-1)^{\lambda'}$)
exists for $\nu=0$ ($K=0$),
and the RPA transition matrix elements
vanishes if the signature mismatches between the state and the operator.
By using the fact that
$$
    \Tr{K}{Q^{(+)}_{\lambda \nu}} =
   -\Tr{K}{Q^{(-)}_{\lambda \nu}} \times (1 + O(\omega^2_\rot)) ,
        \quad (K, \nu > 0) ,  \eqno(3.13)
$$
due to the phase convention (2.7) and
because of the degeneracy of $r=\pm 1$ RPA modes at $\omega_\rot=0$,
the intrinsic matrix elements
in Eq.(2.14) are given by
$$
\eqalignno{
  \rbraketc{\f}{\hm^{(0)}_{\lambda, -K}}{\i}
    &= \Bigl[ \Tr{K}{Q^{(\sigma)}_{\lambda K}} \Bigr]_0 ,  &(3.14\a) \cr
  \rbraketc{\f}{\hm^{(\pm 1)}_{\lambda, -K \pm 1}}{\i}
    &=\sqrt{\frac{2-\delta_{K0}}{2(1+\delta_{K,\pm 1})}}\,
     \frac{1}{\cJ} \Bigl[
       \frac{d\Tr{K}{Q^{(\sigma)}_{\lambda |-K \pm 1|}}}{\omega_\rot}
        \Bigr]_0 ,  &(3.14\b) \cr}
$$
where
$$
     \sigma = \left\{
       \eqalign{
          & + \quad {\rm for} \quad \lambda={\rm even} , \cr
          & - \quad {\rm for} \quad \lambda={\rm odd}  . \cr }
          \right.   \eqno(3.14\c)
$$
The GIR is then given by Eq.(3.8) with
$\Ki=K$, $\Kf=0$ and $\dK = -K \le 0$, namely,
$$
\eqalignno{
   \cT(\lambda:K\,\i \to 0\,\f) =
     &\braketb{\Ii K \lambda \,(-K) }{\If 0} \, Q_t \,
      \bigl( 1 + q \, [\If (\If+1) - \Ii (\Ii+1)] \,
          \bigr) \cr
     &+\sqrt{\If(\If+1)} \,\braketb{\Ii K \lambda \,(-K-1) }{\If \,(-1)} \,
          Q'_t ,  &(3.15) \cr}
$$
and using $C_{\i\f}=\sqrt{2-\delta_{K0}}$ for $\Ki=K$ and $\Kf=0$,
$$
\eqalignno{
   Q_t =&\, \sqrt{2-\delta_{K0}} \,\Bigl[
       \Tr{K}{Q^{(\sigma)}_{\lambda K}} \Bigr]_0
    - \frac{2-\delta_{K0}}{\sqrt{2}}
      \frac{\sqrt{(\lambda-K)(\lambda+K+1)}}{2\cJ} \Bigl[
      \frac{ d\Tr{K}{Q^{(\sigma)}_{\lambda,K+1}} }{d\omega_\rot}
        \Bigr]_0 \cr
     &\quad\quad
    + \frac{2-\delta_{K0}}{\sqrt{2(1+\delta_{K1})}}
      \frac{2K^2+(\lambda-K)(\lambda+K+1)}{\sqrt{(\lambda+K)(\lambda-K+1)}}
         \frac{1}{2\cJ} \Bigl[
      \frac{ d\Tr{K}{Q^{(\sigma)}_{\lambda |K-1|}} }{d\omega_\rot}
    \Bigr]_0 , \cr
   Q'_t =&\, \frac{2-\delta_{K0}}{\sqrt{2}}
        \frac{1}{\cJ} \Bigl[
      \frac{ d\Tr{K}{Q^{(\sigma)}_{\lambda,K+1}} }{d\omega_\rot}
         \Bigr]_0 \cr
     &\quad\quad\quad\quad
    - \frac{2-\delta_{K0}}{\sqrt{2(1+\delta_{K1})}}
        \sqrt{\frac{(\lambda-K)(\lambda+K+1)}{(\lambda+K)(\lambda-K+1)}}\,
         \frac{1}{\cJ} \Bigl[
      \frac{ d\Tr{K}{Q^{(\sigma)}_{\lambda |K-1|}} }{d\omega_\rot}
          \Bigr]_0 , \cr
   q =&\, \frac{2-\delta_{K0}}{\sqrt{2(1+\delta_{K1})}}
         \frac{1}{\sqrt{(\lambda+K)(\lambda-K+1)}}
         \frac{1}{\cJ} \Bigl[
      \frac{ d\Tr{K}{Q^{(\sigma)}_{\lambda |K-1|}} }{d\omega_\rot}
    \Bigr]_0 \, / Q_t .   &(3.16) \cr}
$$
It is easy to see that $Q'_t=0$ for both $K=0$ and $K=\lambda$ cases.
These formula were employed in Ref.\Findref\Ucoul to analyze
the $E1$ and $E3$
transitions from the octupole vibrational bands.

\bigbreak
\vskip 2 mm
\leftline{\it 3-5. $E2$ and $M1$ Transitions
           from quadrupole-vibrational bands}
\vskip 2 mm

     In the previous subsection, we have given the GIR,
Eqs.(3.15) and (3.16), for arbitrary
type of transitions to the ground band in even-even nuclei
($\Kf = 0$), assuming the simple RPA method.
For odd nuclei ($\Kf \ne 0$),
the most important corrections to the RPA may be
taken into account by means of particle-vibration coupling
in the rotating frame\findrefs\MSM\NMM\endrefs.
In principle, the approach presented in \S2 is general and
applicable to such cases.

     Since the occurrence of collective-quadrupole transitions is
a dominant feature of low-lying nuclear spectra,
we now consider the interband $E2$ and $M1$ transitions
from the quadrupole-vibrational bands.
In deformed nuclei, we have the $\beta$ and $\gamma$ vibrations
which transfer the $K$-quantum number by 0 and $\pm 2$ units,
respectively.
In this subsection, we assume that
the band on which the quadrupole vibration is built
has non-zero
$K$-quantum number, and we do not assume the simple RPA method.

     Let us start with the $\beta$ vibration;
the $\beta$ vibration neither transfers the $K$-quantum number
nor changes the signature-quantum number.
The GIR for $E2$ transitions is given by the general
formula in Eq.(3.8) with $\lambda=2$, $\Ki=\Kf=K >0$,
but since $\dK=0$ we must choose either the upper or lower expression;
we here take the upper one.
Then the intrinsic parameters (3.9) are explicitly given
in terms of the matrix elements in the signature-good states as
$$
\eqalignno{
   Q_t =&\, \Bigl[
         \braketc{\f}{Q^{(+)}_{20}}{\i} \Bigr]_0
       +\frac{\sqrt{3}}{\cJ} \Bigl[
        \frac{d\braketc{\bar \f}{Q^{(-)}_{21}}{\i}}{d\omega_\rot}
        \Bigr]_0 , \quad
   Q'_t =\,-\frac{\sqrt{2}}{\cJ} \Bigl[
        \frac{d\braketc{\bar \f}{Q^{(-)}_{21}}{\i}}{d\omega_\rot}
          \Bigr]_0 , \cr
   q =&\, \frac{1}{2\sqrt{3}\cJ} \Bigl[
        \frac{d\braketc{\f}{Q^{(+)}_{21}}{\i}}{d\omega_\rot}
       +\frac{d\braketc{\bar \f}{Q^{(-)}_{21}}{\i}}{d\omega_\rot}
          \Bigr]_0 \, / Q_t , \quad (\dK=0).  &(3.17) \cr}
$$
Here we have used the identities between the matrix elements
shown in Appendix B.
In the simple RPA method, where
$ \ket{\i} = X_\beta^\dagger \ket{\f} $ and $ X_\beta^\dagger $
is the creation operator of the $\beta$ vibration,
the matrix elements of the signature $(-)$ operator,
i.e. $Q^{(-)}_{21}$ term in Eq.(3.17), vanishes.

   Although the interband $M1$ transitions are not necessarily enhanced by
the quadrupole collectivity, the GIR for them can be obtained
in the same way.
The intrinsic parameters in this case are,
$$
\eqalignno{
   Q_t =&\, \Bigl[
         \braketc{\bar \f}{\mu_z}{\i} \Bigr]_0
       -\frac{1}{\cJ} \Bigl[
        \frac{d\braketc{\f}{\mu_x}{\i}}{d\omega_\rot}
        \Bigr]_0 , \quad
   Q'_t =\,\frac{\sqrt{2}}{\cJ} \Bigl[
        \frac{d\braketc{\f}{\mu_x}{\i}}{d\omega_\rot}
          \Bigr]_0 , \cr
   q =& - \frac{1}{2\cJ} \Bigl[
        \frac{d\braketc{\bar \f}{i\mu_y}{\i}}{d\omega_\rot}
       +\frac{d\braketc{\f}{\mu_x}{\i}}{d\omega_\rot}
          \Bigr]_0 \, / Q_t , \quad (\dK=0).  &(3.18) \cr}
$$
Again, $\mu_z$ and $i\mu_y$ terms in Eq.(3.18) vanish
if we use the simple RPA method.
Note that the relative phase between Eqs.(3.17) and (3.18) is meaningful
so that the sign of $M1/E2$ mixing ratio is also determined.

    Now let us consider the case of the $\gamma$ vibration;
the $\gamma$ vibration transfers the $K$-quantum number by $\dK=\pm 2$,
and two independent modes exist, whose signatures are $\alpha=0$ and 1.
Therefore, if $\Kf \ne 0 $, there are two (strongly-coupled) bands
with $\Ki=\Kf +2$ and $\Kf -2$ for a given final state
on which the $\gamma$-vibrational modes are built.
Assuming that both initial and final states
have positive $K$-quantum number at $\omega_\rot=0$
i.e. $\Kf > 2$ for simplicity,
we obtain the intrinsic parameters for $E2$ transitions
between the $\gamma$ band and the ground-state band;
$$
\eqalignno{
   Q_t =& \sqrt{2} \Bigl[
      \braketc{\f}{Q^{(+)}_{22}}{\i}
        \Bigr]_0
     \pm \frac{\sqrt{2}\,(\Ki+\Kf)}{\cJ} \Bigl[
       \frac{d\braketc{\f}{Q^{(+)}_{21}}{\i} }{d\omega_\rot}
         \Bigr]_0 , \quad Q'_t=0, \cr
   q =& \frac{1}{\sqrt{2}\cJ} \Bigl[
       \frac{d\braketc{\f}{Q^{(+)}_{21}}{\i} }{d\omega_\rot}
    \Bigr]_0 \,/Q_t , \quad (\dK= \mp 2) , &(3.19) \cr}
$$
which enter in the GIR (3.8), or equivalently in (3.10) with $n=0$.
Here it is understood that $\ket{\f} $ and $\ket{\i} $
have the same signature and the identities in Appendix B are used.

   The interband $M1$ transitions from the $\gamma$ band is $K$-forbidden
by order $n=1$, and the GIR can be obtained in the same way
as for the $E2$ transitions.
Again, assuming both initial and final states
have non-zero $K$-quantum number at $\omega_\rot=0$,
the GIR is given by Eq.(3.10) with the intrinsic parameters
$$
\eqalignno{
   Q_t =& -\frac{\sqrt{2}}{\cJ} \Bigl[
        \frac{d\braketc{\bar \f}{i\mu_y}{\i}}{d\omega_\rot}
          \Bigr]_0
     \pm \frac{\Ki+\Kf}{2\sqrt{2}\cJ^2} \Bigl[
        \frac{d^2\braketc{\bar \f}{\mu_z}{\i}}{d\omega^2_\rot}
          \Bigr]_0 , \cr
   q =& \frac{1}{2\sqrt{2}\cJ^2} \Bigl[
      \frac{d^2\braketc{\bar \f}{\mu_z}{\i}}{d\omega^2_\rot}
         \Bigr]_0 \,/Q_t , \quad (\dK= \mp 2) . &(3.20) \cr}
$$
Again, the relative sign between Eqs.(3.19) and (3.20) is
related to the sign of the $M1/E2$ mixing ratio and is
determined by the phase convention (2.7).

    Before ending this section, let us discuss
the $M1$ and $E2$ transitions from $\beta$ and $\gamma$ bands
to the ground-state band of even-even
nuclei ($\Kf=0$).
A large body of experimental data exists
and the amplitudes are relatively easy to calculate
by means of the simple RPA method.
Although the general results have been already given
in \S3-4 for the $K$-allowed vibrational transitions, here we
summarize the formulae for the specific case under consideration.
Using the RPA transition amplitudes, $\Tr{\beta}{*}$ and $\Tr{\gamma}{*}$,
for the $\beta$ and $\gamma$ vibrations, respectively,
it is straightforward to show that the formula can be unified into
$$
\eqalignno{
   \cT(E2:K\,\i \to 0\f)
     = &\braketb{\Ii K 2 \,(-K) }{\If 0} \, Q_t \, \cr
       &\quad\times \bigl( 1 + q \, [\If (\If+1) - \Ii (\Ii+1)] \,
         \bigr) ,   &(3.21) \cr
  \cT(M1:K\,\i \to 0\f)
     = &\sqrt{\If(\If+1)}\,
         \braketb{\Ii K 1 \,(1-K)}{\If 1} \, M_t \, \cr
       &\quad\times \bigl( 1 + m \, [\If (\If+1) - \Ii (\Ii+1)] \,
         \bigr) ,   &(3.22) \cr}
$$
with
$$
    Q_t^{(\beta)} = \Bigl[ \Tr{\beta}{Q^{(+)}_{20}} \Bigr]_0 , \quad
    q^{(\beta)} \equiv \frac{1}{2\sqrt{3}\cJ} \Bigl[
       \frac{ d\Tr{\beta}{Q^{(+)}_{21}} }{d\omega_\rot}
        \Bigr]_0 \,/Q_t^{(\beta)} ,  \eqno(3.23\a)
$$
$$
    M_t^{(\beta)} = \frac{\sqrt{2}}{\cJ} \Bigl[
       \frac{ d\Tr{\beta}{\mu_x} }{d\omega_\rot}
        \Bigr]_0 , \quad
    m^{(\beta)}=0 ,    \eqno(3.23\b)
$$
$$
    Q_t^{(\gamma)} =
       \sqrt{2} \Bigl[ \Tr{\gamma}{Q^{(-)}_{22}} \Bigr]_0
         + \frac{2\sqrt{2}}{\cJ} \Bigl[
     \frac{ d\Tr{\gamma}{Q^{(-)}_{21}} }{d\omega_\rot}
        \Bigr]_0 , \quad
    q^{(\gamma)} = \frac{1}{\sqrt{2}\cJ} \Bigl[
       \frac{ d\Tr{\gamma}{Q^{(-)}_{21}} }{d\omega_\rot}
        \Bigr]_0 \,/Q_t^{(\gamma)} ,  \eqno(3.24\a)
$$
$$
    M_t^{(\gamma)} = \frac{\sqrt{2}}{\cJ} \Bigl[
       \frac{ d\Tr{\gamma}{i\mu_y} }{d\omega_\rot}
        \Bigr]_0 - \frac{1}{\sqrt{2}\cJ^2}  \Bigl[
       \frac{ d^2\Tr{\gamma}{\mu_z} }{d\omega^2_\rot}
        \Bigr]_0 , \quad
    m^{(\gamma)}  = - \frac{1}{2\sqrt{2}\cJ^2}  \Bigl[
       \frac{ d^2\Tr{\gamma}{\mu_z} }{d\omega^2_\rot}
        \Bigr]_0 \,/M_t^{(\gamma)} ,  \eqno(3.24\b)
$$
where $\Ki \equiv K=0$ (2) for the $\beta$ ($\gamma$) vibrations,
respectively,
and we have used Eq.(3.13) so as to
make the sign of the $M1/E2$ mixing ratio apparent
(although the overall sign is arbitrary chosen).
These formulae have also been applied to experimental data
in Ref.\Findref\Ucoul.
It should be mentioned that
the GIR (3.21) for the $\gamma$-vibrational band
was derived in Ref.\Findref\SMw,
but because the operator ordering was not considered
the second term in $Q_t^{(\gamma)}$ did not appear.

\bigbreak
\vskip 2 mm
\leftline{ \bf \S4 Examples of Numerical Calculations}
\vskip 2 mm

     In order to show the usefulness of our new approach to the
intensity relationship presented
in the previous sections, we have performed numerical calculations
for transitions from quadrupole- and
octupole-vibrational bands to the ground-state band.
Since the results for $^{238}$U
were already reported in Ref.\Findref\Ucoul,
we have selected even-even nuclei in the rare-earth region,
$^{156,158,160}$Gd, $^{160,162,164}$Dy, and
$^{166,168,170}$Er, for which comprehensive data
have been reported\findrefs\GdDyDT\ErDT\endrefs
for the transition probabilities from both quadrupole- and
octupole-vibrational bands.

    The RPA calculations for the collective vibrational bands
were performed with the cranked-Nilsson mean field
(see the previous works\findrefs\SMa\SMw\MiSM\NMMS\endrefs for
details).\footnote{*}{
    The modified treatment of
    the cranking term\findref\NMMS, which eliminates
    the spurious velocity dependence of the mean field,
    is not used in this paper.}
The standard Nilsson potential with single-stretched
{\twelvebi ll} and {\twelvebi ls} terms\findref\BRnil
was adopted, and we used the experimental quadrupole
deformations\findref\DefDT
which were deduced from the intrinsic $Q$-moments in the ground bands.
The monopole pairing,
quadrupole, and octupole separable forces were employed
as residual interactions.
The force strengths of these interactions was determined at zero
rotational frequency:
The experimental even-odd mass difference, which was calculated by
the third-order-difference formula applied to
the 1993 Mass Table\findref\GapDT,
was used to fix the monopole-pairing strength.
The quadrupole and octupole force strengths for each $K$ value
were chosen so as to reproduce
the observed band-head energies of the corresponding
collective vibrational bands.
The requirement of a zero-energy solution (Nambu-Goldstone mode) was used
for the $K=1$ component of the quadrupole interaction.

In order to estimate the lowest-order corrections of
the Coriolis coupling,
we calculated the RPA transition amplitudes
at finite (infinitesimal) frequency.
The pairing correlations were selfconsistently calculated at
finite frequency,
while the deformation parameters were assumed to be constant.
In principle, the moments of inertia, which are necessary for
calculating the lowest-order corrections,
could be obtained in the cranking model.
However, it is well-known
that the Nilsson model without higher-multipole pairing correlations
generally underestimates moments of inertia near the ground state.
Thus, we use the experimental values estimated
from excitation energies of the first $2^+$ states in the ground band.
The parameters used in the calculation are summarized
in Table 1
(the band-head energies of octupole bands are listed in Table 3).

The transition amplitudes are often
overestimated in RPA calculations with the full model space.
We have used three major shells as the active
model space for both neutrons and protons, i.e.
$N_\osc=4-6$ for neutrons and $N_\osc=3-5$ for protons,
which gives good overall agreement\findref\SMw
for absolute values of transition amplitudes.
Although these absolute values depend on the size of model space,
the quantities representing effects of the Coriolis coupling
($q$ and $m$ in Eqs.(3.15), (3.21), and (3.22))
are not sensitive to it.

    Table 2 shows the calculated results
for the GIR parameters in Eqs.(3.21--24),
associated with the transitions from
the $\beta$ ($K=0$) and $\gamma$ ($K=2$)
vibrational bands,
where they are compared with experimental data
for $I^\pi_\i = 2^+$ states.
The agreement of the $Q_t$ parameters for
the $\gamma$ vibrations is excellent,
while the experimental values for the $\beta$ vibrations
are smaller by about factor two (except in $^{156}$Gd).
The reduction and the fragmentation of measured collectivity
of $\beta$ vibrations
compared with RPA calculations is well-known
in this region\findrefs\GdDyDT\ErDT\endrefs,
and calls for deeper understanding of the low-lying
quadrupole-vibrational modes.
The lowest-order corrections of the Coriolis coupling are generally small
for $E2$ amplitudes: The calculated values of the parameter $q$
are always positive and typically $0.02-0.03$,
which qualitatively agrees with the experimental observations.
Again the agreement
is better for the $\gamma$ band than for the $\beta$ band.

    The intrinsic parameters $Q_t$ and $q$ for the $\gamma$-vibrational
band in some nuclei were calculated previously in Ref.\Findref\SMw.
The differences in those results compared with
the present calculations are due to
the fact that the operator ordering was not considered previously.
As is clear from Eq.(3.24), the extra contribution coming
from the ``contraction'' due to the ordering (2.12) can be written as
$4\,Q^{(\gamma)}_t q^{(\gamma)}$, which increases the leading-order
amplitude $Q^{(\gamma)}_t$
by about 5-10\% and consequently reduces $q^{(\gamma)}$
in comparison with the old results.
In Ref.\Findref\SMw, it was also demonstrated that the ambiguity
of the moment of inertia results in an uncertainty of about 10\%
in the $q$ parameters.

    The $M1$ transitions from quadrupole-vibrational bands to
the ground-state bands are
expected to be small because the coherence property of
the quadrupole phonon is different from that of the $M1$ operator.
This comes about because the $M1$ operator
has the opposite sign to the $E2$ operator
with respect to $H$-conjugation\findref\BMb.
That the $M1$ transitions is weak is apparent
in both the calculations and the experiment.
In particular the $E2/M1$ mixing ratios are large, being typically
$|\delta|>10 $ for the $\gamma$ band\findrefs\GdDyDT\ErDT\endrefs.
Although the $M1$ amplitudes are small, the agreement between
theory and experiment for the
absolute values of the main amplitude parameters $M_t$ is good.
It should be noted that, experimentally,
the sign of the mixing amplitude
is negative for all the $\gamma$ bands in the table,
while the calculation indicates that the sign changes from negative
to positive as one moves up to heavier isotopes.
It is worthwhile mentioning that the $M1$ transitions
from the quadrupole-vibrational bands to the yrast bands
are predicted to become large at high spin\findref\SMw.
This is because the strong Coriolis coupling changes
the nature of the quadrupole vibration into a wobbling-like motion
and the coherence property of the phonon begins to
favor the $M1$ operator.
The properties of $M1$ transitions from high-spin vibrational excitations
is an interesting problem for the future.

    In Table 3, we show the calculated results
for the octupole-vibrational bands (see Eq.(3.16) with
$\lambda=3$).  Since there are not enough data to extract
the GIR parameters, we have compared the calculated and
experimental amplitudes of the stretched vibrational transitions,
$I^\pi=3^-_K \to 0^+_\g$.
It has been discussed in Ref.\Findref\NV that Coriolis
mixing amongst the octupole vibrations
with different $K$
is important even at low spins.
The measured relative intensities of transitions associated with
the different $K$ modes cannot be understood
without Coriolis-coupling effects.
The reason why Coriolis coupling for the $E3$
transitions ($3^-_K \rightarrow 0^+_\g$) is more important than for
the $E2$ ($2^+_{\beta,\gamma} \rightarrow 0^+_\g$)
can be understood as follows:
In principle, all members of the octupole multiplet with $K=0$, 1, 2, and 3
can occur as low-energy modes of excitation,
but there is no low-lying quadrupole vibration with $K=1$.
Therefore, the octupole bands are expected to be mixed
with each other more strongly than are the quadrupole bands ($\dK = 2$).
In addition,
for the stretched vibrational transitions to the ground state,
the factor multiplying the Coriolis-coupling
parameter $q$ in Eq.(3.15), namely $-\lambda(\lambda+1)$,
takes the value $-12$ for the octupole bands
and $-6$ for the quadrupole bands.
Typically, the lowest-order coupling changes the $E2$ amplitudes
by less than $\sim$10\%, but it can
change the $E3$ amplitudes by more than 50\%.

    Because we are restricting this discussion to the $\If=0$ case,
the second ($Q_t'$) term in Eq.(3.15) does not contribute
to the $E3$ amplitudes.
However, the values of parameter $Q_t'$ in Table 3 provide
qualitative information concerning the magnitude of the Coriolis mixing.
Since our treatment is perturbative in its nature,
the results will not be reliable if the Coriolis coupling is too strong.
A naive criterion for the validity of our approach with respect to
the Coriolis coupling may be given by $|Q_t'| \ll |Q_t|$.
This criterion is broken for some nuclei
in which two octupole modes
with adjacent $K$-quantum numbers are almost degenerate, e.g.
$K=1$ and 2 in $^{160}$Dy,
$K=0$ and 1 in $^{164}$Dy, and $K=2$ and 3 in $^{168}$Er
(there is a similar degeneracy for the
$\beta$ and $\gamma$ bands in $^{158}$Gd).
In those cases, the lowest-order correction
$\rbraketc{\f}{\hm^{(\pm 1)}_{3 \dK\pm 1}}{\i}$
is comparable to the main amplitude
$\rbraketc{\f}{\hm^{(0)}_{3 \dK}}{\i}$,
therefore
the second and third terms in the expression of $Q_t$ (3.16)
will be even larger than the first (leading) term,
indicating a breakdown of perturbation theory.

Table 3 also presents amplitudes calculated without the effect
of the Coriolis coupling
(the column ``cal.(0)'').
As is clear from the table,
the calculated values of the parameter $q$ for
the lowest octupole states are always negative (except for $^{160}$Dy).
This has the effect of concentrating $E3$ strength
onto the lowest state.
This property agrees with the results of Ref.\Findref\NV.
Except for the ``singular'' cases mentioned above,
agreement with the experimental data is significantly
improved by the lowest-order corrections.

\bigbreak
\vskip 2 mm
\leftline{ \bf \S5. Concluding Remarks }
\vskip 2 mm

    In this paper we have developed a new general method
to calculate the intrinsic parameters of the generalized intensity
relations for transition probabilities.
The intrinsic matrix elements in the original formulation
of the unified model\findref\BMb
are related to matrix elements
in the microscopic cranking calculations.
Useful formula
for the electromagnetic transition amplitudes
are presented.

In our approach,
transition probabilities
are microscopically calculated by the cranking model.
Since the geometry of angular momentum
(Clebsch-Gordan coefficients) is fully taken into account,
our results are applicable to low-spin states.
We have presented the numerical results of cranked RPA calculations
giving transition amplitudes
between vibrational and ground-state bands.
These examples
clearly show that our formalism gives an improved
understanding of the transition rates between these
low-lying collective motions.

   The generalized intensity relation is derived
by a perturbative treatment of the intrinsic-angular-momentum
operator ($I$ expansion)\findref\BMb.
In principle,
the correction terms to the leading-order relation
could be evaluated by explicitly calculating the perturbation of
the Coriolis couplings in the unified model.
This may be relatively easy for single-particle states,
but it becomes much more difficult for correlated
states such as collective vibrations.
In contrast, our method gives a systematic and practical means
to account for the Coriolis coupling in terms of the cranking
formalism.

  Until now, the cranking model has been considered unreliable in
calculating transition matrix elements,
especially at low spin.
This is indeed so if the angular momentum is treated
only on average, as in the semiclassical approximation.
In the high-spin limit, where the rotational motion is uniform
around one of the principal axis of the mean field,
the effect of angular-momentum algebra is not important
because the Clebsch-Gordan coefficients can be replaced
by their asymptotic values ($1/I$ expansion)\findref\Marb.
The formalism developed in this paper gives a complementary
method which is a good approximation for low spins,
even though it uses the same apparatus, namely the cranking model.

   Although the idea of the present approach is general,
we have restricted its application to the lowest-order
corrections because the problem of operator ordering
appears to be more complex for the higher orders.
In order to extend the formulation to the higher orders,
a more systematic quantum-mechanical treatment of
the microscopic cranking model may be necessary\findref\BesT.
This is an important future problem.

\bigbreak
\vskip 2 mm
\leftline{ \bf Acknowledgements }
\vskip 2 mm

   This work is support in part by the Grant-in-Aid for
Scientific Research from the Japan Ministry of Education,
Science and Culture (No. 08640383).
We thank the Institute for Nuclear Theory
at the University of Washington for its hospitality
and the U.S. Department of Energy for partial support during
the completion of this work.

\vfill
\eject
\baselineskip 16 truept
\parindent=0 truept
Table 1. \quad
Basic parameters used in the calculations.
The neutron and proton pairing gaps ($\Delta_{\n,\p}$)
are obtained by third
differences from the 1993 Mass Table\findref\GapDT.
The Nilsson-deformation parameters ($\epsilon_2$)
are taken from Ref.\Findref\DefDT.
The moments of inertia ($\cJ_\g$) are obtained from the first
$2^+$ energies of the ground-state bands.
The band-head energies of the $\gamma$ and $\beta$ vibrations
($E_{\gamma,\beta}$)
are taken from the compilation of Ref.\Findref\VibDT.
\vskip -3 truemm
$$
\vbox{
\halign{\strut
 \quad \hfil # \hfil & \quad \hfil # \hfil & \hfil # \hfil &
       \hfil # \hfil & \hfil # \hfil & \hfil # & \hfil # \cr
\noalign{\hrule}
\noalign{\vskip 1 truemm}
  Nucl. & $\Delta_\n\,$[MeV]  &  $\Delta_\p\,$[MeV] & $\epsilon_2$ &
   $\cJ_\g\,$[$\hbar^2$/MeV]$\,\,$ &
   \ $E_\gamma\,$[keV] \hfil & $E_\beta\,$[keV] \hfil \cr
\noalign{\vskip 1 truemm}
\noalign{\hrule}
\noalign{\vskip 1 truemm}
 $^{156}$Gd & 1.070 & 0.960 & 0.274 & 33.7 & 1154\quad & 1050\quad \cr
 $^{158}$Gd & 0.892 & 0.878 & 0.282 & 37.7 & 1187\quad & 1196\quad \cr
 $^{160}$Gd & 0.831 & 0.871 & 0.287 & 39.9 &  988\quad & 1378\quad \cr
\noalign{\vskip 1 truemm}
 $^{160}$Dy & 0.967 & 0.978 & 0.271 & 34.6 &  966\quad & 1275\quad \cr
 $^{162}$Dy & 0.917 & 0.930 & 0.270 & 37.2 &  888\quad & 1131\quad \cr
 $^{164}$Dy & 0.832 & 0.875 & 0.275 & 40.9 &  762\quad & 1655\quad \cr
\noalign{\vskip 1 truemm}
 $^{166}$Er & 0.966 & 0.877 & 0.272 & 37.2 &  786\quad & 1460\quad \cr
 $^{168}$Er & 0.776 & 0.857 & 0.271 & 37.6 &  821\quad & 1217\quad \cr
 $^{170}$Er & 0.708 & 0.797 & 0.268 & 38.2 &  934\quad &  891\quad \cr
\noalign{\vskip 1 truemm}
\noalign{\hrule}
}}
$$
\vfill
\eject

\baselineskip 16 truept
\parindent=0 truept
Table 2. \quad
Comparison of the GIR parameters for $E2$ and $M1$ transitions
[Eqs.(3.23) and (3.24)]
from the $\gamma\,(K=2)$ and $\beta\,(K=0)$ bands to the ground-state band.
Experimental values\findrefs\GdDyDT\ErDT\endrefs
for $Q_t$ ($M_t$) are extracted from
$B(E2(M1); 2^+_{\beta,\gamma} \to 2^+_\g)$
and those for $q$ are extracted both from
$B(E2; 2^+_{\beta,\gamma} \to 0^+_\g)$ (indicated by ``exp.(1)'')
and $B(E2; 2^+_{\beta,\gamma} \to 4^+_\g)$ (``exp.(2)'').
Note that the relative phase between $E2$ and $M1$ matrix elements
(the $E2/M1$ mixing ratio) is meaningful
in both the experiment and the calculation.
\vskip -3 truemm
$$
\vbox{
\halign{\strut
 \quad \hfil # \hfil & \hfil # \hfil
        & \quad \hfil # \hfil & \hfil # \hfil  
        & \quad \hfil # & \hfil # & \hfil #          
        & \quad \hfil # & \hfil #                    
        & \quad \hfil #  \cr
\noalign{\hrule}
\noalign{\vskip 1 truemm}
  Nucl. &   & \multispan2 \hfil $Q_t\,$[eb]\hfil
        & \multispan3 \hfil $q$ \hfil
        & \multispan2 \hfil $M_t\,[\mu_\N]$ \hfil
        & $m$ \quad\cr
        &   & cal. & exp.
        & cal. & exp.(1) & exp.(2)
        & cal. & exp.
        & cal. \cr
\noalign{\vskip 1 truemm}
\noalign{\hrule}
\noalign{\vskip 1 truemm}
 $^{156}$Gd & $\gamma$ & 0.391 & 0.352 & 0.024 & 0.009 & 0.025
 & $-$0.0059 & $-$0.0059 & 0.078 \cr
\noalign{\vskip -1 truemm}
            & $\beta$  & 0.171 & 0.267 & 0.042 & 0.088 & $-$0.023
 & $-$0.0115 & $-$0.0066 & \cr
 $^{158}$Gd & $\gamma$ & 0.329 & 0.333 & 0.020 & 0.021 & $-$0.012
 & $-$0.0026 & $-$0.0128 & 0.120 \cr
\noalign{\vskip -1 truemm}
            & $\beta$  & 0.132 & 0.066 & 0.015 & $-$0.061 & 0.057
 & 0.0448 & $-$0.0191 & \cr
 $^{160}$Gd & $\gamma$ & 0.328 & 0.323 & 0.021 & 0.014 & 0.0026
 & 0.0008 & $-$0.0009\rlap{$^*$} & $-$0.375 \cr
\noalign{\vskip -1 truemm}
            & $\beta$  & 0.116 &       & 0.012 &       &
 & 0.0515 &          & \cr
 $^{160}$Dy & $\gamma$ & 0.412 & 0.395 & 0.025 & 0.019 & 0.012
 & $-$0.0037 & $-0.0035$ & 0.128 \cr
\noalign{\vskip -1 truemm}
            & $\beta$  & 0.160 & 0.114 & 0.022 & $-$0.033 & 0.011
 & 0.0241 & $-$0.0245 & \cr
 $^{162}$Dy & $\gamma$ & 0.397 & 0.395 & 0.024 & 0.022 & 0.032
 & 0.0002 & $-$0.0011\rlap{$^*$} & $-$2.890 \cr
\noalign{\vskip -1 truemm}
            & $\beta$  & 0.226 &       & 0.021 &       &
 & 0.0333 &    & \cr
 $^{164}$Dy & $\gamma$ & 0.412 & 0.379 & 0.021 & 0.019 & 0.035
 & 0.0070 & $-$0.0018 & $-$0.050 \cr
\noalign{\vskip -1 truemm}
       & $\beta^\dagger$ & 0.010 & & 0.624 & &
 &$-$0.1588 & & \cr
 $^{166}$Er & $\gamma$ & 0.442 & 0.436 & 0.022 & 0.024 & 0.030
 & 0.0019 & $-$0.0039 & $-$0.227 \cr
\noalign{\vskip -1 truemm}
            & $\beta$  & 0.193 &       & 0.031 & &
 &$-$0.0370 & & \cr
 $^{168}$Er & $\gamma$ & 0.439 & 0.419 & 0.016 & 0.020 & 0.007
 & 0.0094 & $-$0.0039\rlap{$^*$} & $-$0.031 \cr
\noalign{\vskip -1 truemm}
            & $\beta$ & 0.158 &       & 0.037  & &
 &$-$0.0367 & &  \cr
 $^{170}$Er & $\gamma$ & 0.421 & 0.363 & 0.013 & 0.020 & 0.004
 & 0.0119 & $-$0.0014\rlap{$^*$} & $-$0.018 \cr
\noalign{\vskip -1 truemm}
            & $\beta$  & 0.208 & 0.099 & 0.023 & 0.016 & 0.035
 & 0.0147 & 0.0129 &  \cr
\noalign{\vskip 1 truemm}
\noalign{\hrule}
}}
$$
\vskip -5 truemm
\hskip 1 truecm $^{*)}$ {\tenrm
  Only a lower limit for the absolute value is given.}
\vskip -1.5 truemm
\hskip 1 truecm $^{\dagger)}$ {\tenrm
  An almost pure two quasiparticle state
  in the calculation using the energy in Table 1.}
\vfill
\eject

\baselineskip 16 truept
\parindent=0 truept
Table 3. \quad  Calculated GIR parameters ($Q_t$, $Q_t'$, $q$)
and $E3$ amplitudes (cal.(0), cal.) in Eq.(3.15)
for $3^- \to 0^+_\g$ transitions from the octupole-vibrational bands.
``cal.(0)'' indicates the calculated $E3$ amplitudes neglecting the
Coriolis coupling.
Observed band-head energies\findref\VibDT ($E_K$) and
$E3$ amplitudes\findrefs\GdDyDT\ErDT\endrefs (exp.) are also
listed.
\vskip -5 truemm
$$
\vbox{
\offinterlineskip
\halign{\strut
 \quad \hfil # \hfil & \quad
       \hfil # \hfil & \hfil # & \hfil # & \hfil # &
       \hfil # & \hfil # & \hfil # & \hfil #  \cr
\noalign{\hrule}
\noalign{\vskip 1 truemm}
  Nucl. & $K$ & $E_K$\hfil & $Q_t\,$[eb$^\frac{3}{2}$]\hfil &
  $Q'_t\,$[eb$^\frac{3}{2}$]\hfil & $q$\hfil &
  \multispan3\hfil$
 \cT(E3:3^- \to 0^+_\g)\,$[eb$^\frac{3}{2}$]\quad\hfill \cr
  &&& cal. & cal. & cal. &
  \quad cal.(0)\hfil &$\,\,$ cal.\quad \hfil & exp.\hfil \cr
\noalign{\vskip 1 truemm}
\noalign{\hrule}
\noalign{\vskip 1 truemm}
 $^{156}$Gd
  & 0 & 1367 & 0.161 & 0\hfil  & 0.116 &  0.061 & 0.024 & 0.043\rlap{$^*$} \cr
  & 1 & 1243 & 0.111 & 0.065 &$-$0.227 &  0.090 & 0.156 & 0.156 \cr
  & 2 & 1780 & 0.284 &$-$0.057 & 0.022 &  0.071 & 0.079 & \cr
  & 3 & 1934 & 0.342 & 0\hfil & 0.042 &  0.081 & 0.065 & \cr
\noalign{\vskip 1 truemm}
 $^{158}$Gd
  & 0 & 1264 & 0.150 & 0\hfil & 0.045 &  0.057 & 0.026 & 0.018 \cr
  & 1 &  977 & 0.210 & 0.020 &$-$0.040 &  0.094 & 0.117 & 0.133 \cr
  & 2 & 1794 & 0.199 &$-$0.031 & 0.004 &  0.060 & 0.072 & \cr
  & 3 &      & 0.130 & 0\hfil & 0.050 &  0.027 & 0.019 & \cr
\noalign{\vskip 1 truemm}
 $^{160}$Gd
  & 0 & 1225 & 0.146 & 0\hfil &$-$0.019 &  0.055 & 0.067 & 0.130 \cr
  & 1 & 1569 & 0.139 & 0.008 & 0.027 &  0.056 & 0.036 & \cr
  & 2 & 1016 & 0.281 &$-$0.008 &$-$0.011 &  0.107 & 0.120 & \cr
  & 3 & 1463 & 0.314 & 0\hfil & 0.023 &  0.094 & 0.086 & \cr
\noalign{\vskip 2 truemm}
 $^{160}$Dy
  & 0 &      & 0.063 & 0\hfil & 0.063 &  0.024 & 0.006 & \cr
  & 1 & 1286 & 0.533 &$-$0.474 & 0.005 &  0.083 & 0.190 & 0.096%
                                                      \rlap{$^{\dagger}$} \cr
  & 2 & 1265 & 0.619 &$-$0.302 & 0.204 &  0.097 & 0.340 & 0.156 \cr
  & 3 & 2075 & 0.224 & 0\hfil & 0.018 &  0.071 & 0.066 & \cr
\noalign{\vskip 1 truemm}
 $^{162}$Dy
  & 0 & 1276 & 0.157 & 0\hfil &$-$0.023 &  0.059 & 0.076 & 0.070%
                                                      \rlap{$^\ddagger$} \cr
  & 1 & 1637 & 0.175 & 0.002 & 0.034 &  0.065 & 0.039 & \cr
  & 2 & 1148 & 0.260 &$-$0.009 &$-$0.016 &  0.101 & 0.118 & 0.122 \cr
  & 3 & 1571 & 0.324 & 0\hfil & 0.026 &  0.094 & 0.085 & \cr
\noalign{\vskip 1 truemm}
 $^{164}$Dy
  & 0 & 1675 & 0.120 & 0\hfil & 0.294 &  0.045 & 0.115 & \cr
  & 1 & 1637 & 0.072 &$-$0.147 & 0.587 &  0.077 & 0.165 & \cr
  & 2 &  977 & 0.277 &$-$0.0001 &$-$0.011 &  0.109 & 0.118 & 0.112 \cr
  & 3 & 1766 & 0.210 & 0\hfil & 0.021 &  0.065 & 0.060 & \cr
\noalign{\vskip 1 truemm}
\noalign{\hrule}
}}
$$
\vfill
\eject

Table 3. ({\it continued}).
\vskip -5 truemm
$$
\vbox{
\offinterlineskip
\halign{\strut
 \quad \hfil # \hfil & \quad
       \hfil # \hfil & \hfil # & \hfil # & \hfil # &
       \hfil # & \hfil # & \hfil # & \hfil #  \cr
\noalign{\hrule}
\noalign{\vskip 1 truemm}
  Nucl. & $K$ & $E_K$\hfil & $Q_t\,$[eb$^\frac{3}{2}$]\hfil &
  $Q'_t\,$[eb$^\frac{3}{2}$]\hfil & $q$\hfil &
  \multispan3\hfil$
 \cT(E3:3^- \to 0^+_\g)\,$[eb$^\frac{3}{2}$]\quad\hfill \cr
  &&& cal. & cal. & cal. &
  \quad cal.(0)\hfil &$\,\,$ cal.\quad \hfil & exp.\hfil \cr
\noalign{\vskip 1 truemm}
\noalign{\hrule}
\noalign{\vskip 1 truemm}
 $^{166}$Er
  & 0 & 1663 & 0.143 & 0\hfil &$-$0.065 &  0.054 & 0.096 & 0.068 \cr
  & 1 & 1831 & 0.251 &$-$0.024  & 0.051 &  0.076 & 0.036 & \cr
  & 2 & 1458 & 0.228 &$-$0.006  &$-$0.022 &  0.091 & 0.109 & 0.093 \cr
  & 3 & 1918 & 0.196 & 0\hfil & 0.033 &  0.053 & 0.045 & \cr
\noalign{\vskip 1 truemm}
 $^{168}$Er
  & 0 & 1786 & 0.104 & 0\hfil & 0.031 & 0.039 & 0.025 & \cr
  & 1 & 1359 & 0.191 &$-$0.008  &$-$0.020 & 0.069 & 0.089 & 0.078 \cr
  & 2 & 1569 & 0.031 &$-$0.211  &$-$0.106 & 0.081 & 0.027 & 0.085 \cr
  & 3 & 1542 & 0.642 & 0\hfil & 0.147 & 0.079 & 0.186 & \cr
\noalign{\vskip 1 truemm}
 $^{170}$Er
  & 0 & 1539 & 0.051 & 0\hfil & 0.008 & 0.019 & 0.017 & \cr
  & 1 & 1266 & 0.074 & 0.002 &$-$0.028 & 0.030 & 0.037 & \cr
  & 2 &      & 0.124 & 0.023 &$-$0.005 & 0.059 & 0.049 & \cr
  & 3 & 1304 & 0.221 & 0\hfil &$-$0.017 & 0.096 & 0.100 & \cr
\noalign{\vskip 1 truemm}
\noalign{\hrule}
}}
$$
\vskip -5 truemm
\hskip 2 truecm $^{*)}$ {\tenrm
   Only the upper limit is given.\par}
\vskip -1.5 truemm
\hskip 2 truecm $^{\dagger)}$ {\tenrm
   Data from Ref.\Findref\NDSRe.}
\vskip -1.5 truemm
\hskip 2 truecm $^{\ddagger)}$ {\tenrm
   Data from Ref.\Findref\NDSHe.}
\vfill
\eject


\parindent=20 truept
\baselineskip 22 truept

\bigbreak
\vskip 2 mm
\centerline{ \bf Appendix A}
\vskip 2 mm

   In this appendix we discuss the possible ordering of operators,
$I_\pm$ and $\cD^\lambda_{\mu \nu}$,
in the ``quantization'' procedure,
and show that the symmetrized ordering
adopted in Eq.(2.12b) is the most appropriate.

We consider here the simple case where
the $\lambda$-pole transitions occur
from the $\lambda$-pole vibrational states ($\Ki=K$, $\Ii=\lambda$)
to the ground state ($\Kf=\If=0$).
The Coriolis coupling is assumed to be effective only within
the $\lambda$-pole vibrational multiplets
with different $K$-quantum numbers
($K=-\lambda, -\lambda + 1, \cdots, \lambda$).
We impose the following two conditions;
(i) detailed balance is satisfied, i.e.,
$$
    |\braketr{00}{\cM(\lambda)}{K \lambda}|
    = |\braketr{K \lambda}{\cM(\lambda)}{00}| ,  \eqno(\A.1)
$$
and (ii) the sum of the transition strengths for the modes
with different $K$ values is preserved within the first order
in $\omega_\rot$, i.e.,
$$
    \sum_{K=-\lambda}^\lambda
       | \braketr{00}{\cM(\lambda)}{K \lambda}_\unsym |^2
    = \Bigl[ \sum_{K=-\lambda}^\lambda
       | \braketr{00}{\cM(\lambda)}{K \lambda}_\unsym |^2 \Bigr]_0
          + O(\omega_\rot^2) ,          \eqno(\A.2)
$$
where $\braketr{}{*}{}_\unsym$ means that
the $K$-good representation for
the wave functions is used, i.e., the symmetrization according to
$\cR$ conjugation is not considered.
After the summation with respect to $K$ in Eq.(A.2),
this treatment gives the same result as the case where
the $\cR$ conjugation is taken into account.
The symbol $[*]_0$ means that
it is obtained neglecting the Coriolis coupling ($\omega_\rot=0$).

     From the first requirement (A.1), it is easy to see
that only the following orderings are allowed:
$$
\eqalign{
    &\cM(\lambda \mu)_{\dK} \cr
    &\quad= \left\{
   \eqalign{
     &\hm^{(0)}_{\lambda \Delta K} \, \cD^\lambda_{\mu \Delta K}
    + \hm^{(+1)}_{\lambda \Delta K+1}
           \frac{1}{2} \bigl\{ I_+,\cD^\lambda_{\mu, \Delta K+1} \bigr\}
    + \hm^{(-1)}_{\lambda \Delta K-1} \,
           \frac{1}{2} \bigl\{ I_-,\cD^\lambda_{\mu, \Delta K-1} \bigr\} ,
                \hskip  8.3 truemm (\A.3\a) \cr
     &\hm^{(0)}_{\lambda \Delta K} \, \cD^\lambda_{\mu \Delta K}
    + \hm^{(+1)}_{\lambda \Delta K+1} \,
        I_+ \cD^\lambda_{\mu, \Delta K+1}
    + \hm^{(-1)}_{\lambda \Delta K-1} \,
        \cD^\lambda_{\mu, \Delta K-1} I_- ,
                \hskip 27.0 truemm (\A.3\b) \cr
     &\hm^{(0)}_{\lambda \Delta K} \, \cD^\lambda_{\mu \Delta K}
    + \hm^{(+1)}_{\lambda \Delta K+1} \,
        \cD^\lambda_{\mu, \Delta K+1} I_+
    + \hm^{(-1)}_{\lambda \Delta K-1} \,
        I_- \cD^\lambda_{\mu, \Delta K-1} .
                \hskip 28.0 truemm (\A.3\c) \cr}
     \right.  }
$$
We have adopted the ordering of Eq.(A.3a) in the ``quantization'' (2.12b).
The ordering given in Eq.(A.3b) or (A.3c) leads a different result
for the GIR.

Next, in order to check the second requirement (A.2),
we calculate the effect of the Coriolis coupling
in the lowest order of $\omega_\rot$.
Using Eq.(2.14), we obtain
$$
\eqalign{
   &\sum_{K=-\lambda}^\lambda
      | \braketr{00}{\cM(\lambda)}{K \lambda}_\unsym |^2
    - \Bigl[ \sum_{K=-\lambda}^\lambda
      | \braketr{00}{\cM(\lambda)}{K \lambda}_\unsym |^2 \Bigr]_0 \cr
   &\quad \approx \left\{
   \eqalign{
    &-\frac{1}{\cJ} \sum_{K=-\lambda}^\lambda
        \sqrt{(\lambda-K)(\lambda+K+1)}     \cr
    &\quad\quad \times  \Bigl[
      \frac{d\Tr{K}{Q_{\lambda,-K-1}}}{d\omega_\rot}
         \Tr{K}{Q_{\lambda,-K}}
     +\frac{d\Tr{K+1}{Q_{\lambda,-K}}}{d\omega_\rot}
         \Tr{K+1}{Q_{\lambda,-K-1}} \Bigr]_0 ,
                  \hskip  4.2 truemm (\A.4\a) \cr
    &-\frac{2}{\cJ} \sum_{K=-\lambda}^\lambda
        \sqrt{(\lambda-K)(\lambda+K+1)} \, \Bigl[
      \frac{d\Tr{K}{Q_{\lambda,-K-1}}}{d\omega_\rot}
         \Tr{K}{Q_{\lambda,-K}} \Bigr]_0,
                  \hskip 17.0 truemm (\A.4\b) \cr
    &-\frac{2}{\cJ} \sum_{K=-\lambda}^\lambda
        \sqrt{(\lambda-K)(\lambda+K+1)} \, \Bigl[
      \frac{d\Tr{K+1}{Q_{\lambda,-K}}}{d\omega_\rot}
         \Tr{K+1}{Q_{\lambda,-K-1}} \Bigr]_0,
                  \hskip  9.0 truemm (\A.4\c) \cr}
     \right.  }
$$
corresponding to the orderings (A.3a)$-$(A.3c), respectively.
Here the notation of the transition amplitude for the $K$ mode,
$\Tr{K}{*}\equiv \dbraketc{0_\g}{*}{K}$, is used, where $\dket{K}$
denotes the $K$-good states of the $\lambda$-pole vibrational multiplet.

     Since we are considering the Coriolis coupling
only within the $\lambda$-pole multiplets up to the first order
of $\omega_\rot$,
we may write
$$
    \dket{K} = \Bigl[ \dket{K} \Bigr]_0
            + \epsilon_K^{K+1} \Bigl[ \dket{K+1} \Bigr]_0
            + \epsilon_K^{K-1} \Bigl[ \dket{K-1} \Bigr]_0
            + O(\omega_\rot^2) .   \eqno(\A.5)
$$
Then by the orthogonality, $\dbraketb{K+1}{K}=0$, we have
$$
    \epsilon^{K+1}_{K} + \epsilon^K_{K+1} \approx 0 .  \eqno(\A.6)
$$
On the other hand, by the definition of transition amplitude,
$$
    \dbraketc{0_\g}{Q_{\lambda, -K \mp 1}}{K}
     \approx \epsilon_K^{K \pm 1} \, \Bigl[
        \Tr{K \pm 1}{Q_{\lambda, -K \mp 1}} \Bigr]_0
     \approx \omega_\rot \, \Bigl[
        \frac{d\Tr{K}{Q_{\lambda, -K \mp 1}}}{d\omega_\rot} \Bigr]_0,
$$
namely,
$$
   \epsilon_K^{K \pm 1} \approx \omega_\rot \, \Bigl[
      \frac{d\Tr{K}{Q_{\lambda,-K \mp 1}}}{d\omega_\rot}
         /\Tr{K \pm 1}{Q_{\lambda,-K \mp 1}} \Bigr]_0.    \eqno(\A.7)
$$
Combining (A.6) and (A.7), we obtain
$$
       \Bigl[
      \frac{d\Tr{K}{Q_{\lambda,-K-1}}}{d\omega_\rot}
         \Tr{K}{Q_{\lambda,-K}}
     +\frac{d\Tr{K+1}{Q_{\lambda,-K}}}{d\omega_\rot}
         \Tr{K+1}{Q_{\lambda,-K-1}} \Bigr]_0 = 0 .  \eqno(\A.8)
$$
Using this identity for Eq.(A.4),
the requirement (A.2) turns out to be satisfied
for an arbitrary multipole order $\lambda$,
only when the symmetrized ordering (A.3a) is used in Eq.(2.12b).

\bigbreak
\vskip 2 mm
\centerline{ \bf Appendix B}
\vskip 2 mm

    In this appendix we examine the general relation
between the matrix elements in the signature-good basis,
which is used in the cranking calculations,
and those in the $K$ good-representation, which appears in the GIR.

    Let us assume that the initial and final states,
$\ket{\i}$ and $\ket{\f}$, have the same signature and $\Ki,\Kf > 0$.
Using Eq.(2.7b) and (3.1), the non-zero matrix elements of the operator
$\cQ^{(\pm)}_{\lambda\nu}(\nu \ge 0)$ in the signature-good basis are
written in terms of the $K$-good basis as
$$
   \left.
   \eqalign{&\braketc{\f}{Q^{(+)}_{\lambda\nu}}{\i} \cr
            &\braketc{\bar\f}{Q^{(+)}_{\lambda\nu}}{\bar\i} \cr
            &\braketc{\bar\f}{Q^{(-)}_{\lambda\nu}}{\i} \cr
            &\braketc{\f}{Q^{(-)}_{\lambda\nu}}{\bar\i} \cr }
   \right\}= \frac{1}{\sqrt{2(1+\delta_{\nu 0})}} \times
   \left\{
   \eqalign{&( A + B + C + D )\cr
            &( A + B - C - D )\cr
            &( A - B + C - D )\cr
            &( A - B - C + D )\cr }
   \right.       \eqno(\B.1)
$$
with
$$
\eqalignno{
    A &= \frac{1}{2}\Bigr(
         \dbraketc{+\f}{Q_{\lambda\nu}}{+\i} \bigl) +
         (-)^\lambda \dbraketc{-\f}{Q_{\lambda,-\nu}}{-\i} \Bigl)
       = O(\omega_\rot^{|\dK-\nu|}) ,   &(\B.2\a) \cr
    B &= \frac{1}{2}\Bigr(
         \dbraketc{-\f}{Q_{\lambda\nu}}{-\i} \bigl) +
         (-)^\lambda \dbraketc{+\f}{Q_{\lambda,-\nu}}{+\i} \Bigl)
       = O(\omega_\rot^{|\dK+\nu|}) ,   &(\B.2\b) \cr
    C &= \frac{1}{2}\Bigr(
         \dbraketc{+\f}{Q_{\lambda\nu}}{-\i} \bigl) +
         (-)^\lambda \dbraketc{-\f}{Q_{\lambda,-\nu}}{+\i} \Bigl)
       = O(\omega_\rot^{|\Ki+\Kf-\nu|}) ,   &(\B.2\c) \cr
    D &= \frac{1}{2}\Bigr(
         \dbraketc{-\f}{Q_{\lambda\nu}}{+\i} \bigl) +
         (-)^\lambda \dbraketc{+\f}{Q_{\lambda,-\nu}}{-\i} \Bigl)
       = O(\omega_\rot^{|\Ki+\Kf+\nu|}) ,   &(\B.2\d) \cr }
$$
where the notation $\dket{\pm \i}$ in Eq.(2.7b) is used.
Since the $\omega_\rot \to 0$ limit is taken to evaluate
the intrinsic parameters entering the GIR,
not all the amplitudes in Eq.(B.2a--d) contribute.
In the case of $\Ki+\Kf > |\dK|$
under consideration ($\Ki,\Kf >0$),
$C$ and $D$ terms do not contribute;
if in addition $\nu, \dK \ne 0$,
only the $A(B)$ term survives for $\dK >0(<0)$.
By using this property we can easily derive identities
between the matrix elements (or the derivatives of them with respect
to $\omega_\rot$) in the left hand side of Eq.(B.1) at $\omega_\rot=0$.

     For example, the $\dK=0$ $E2$ transitions with $\Ki=\Kf>0$,
we obtain the following identities:
$$
       \Bigl[ \braketc{\bar \f}{Q^{(+)}_{20}}{\bar \i} \Bigr]_0
     = \Bigl[ \braketc{\f}{Q^{(+)}_{20}}{\i} \Bigr]_0,
           \eqno(\B.3\a)
$$
$$
       \Bigl[
        \frac{d\braketc{\bar \f}{Q^{(+)}_{21}}{\bar \i}}
          {d\omega_\rot} \Bigr]_0
     = \Bigl[
        \frac{d\braketc{\f}{Q^{(+)}_{21}}{\i}}
          {d\omega_\rot} \Bigr]_0 , \quad
       \Bigl[
        \frac{d\braketc{\f}{Q^{(-)}_{21}}{\bar \i}}
          {d\omega_\rot} \Bigr]_0
     = \Bigl[
        \frac{d\braketc{\bar \f}{Q^{(-)}_{21}}{\i}}
          {d\omega_\rot} \Bigr]_0 ,  \eqno(\B.3\b)
$$
which together with similar identities for the $M1$ operators
were used to derive Eqs.(3.17) and (3.18) for the transitions
from the $\beta$ band.
For the $\dK =\mp 2$ $E2$ transitions, similar identities,
$$
       \Bigl[ \braketc{\bar \f}{Q^{(+)}_{22}}{\bar \i} \Bigr]_0
     = \Bigl[ \braketc{\f}{Q^{(+)}_{22}}{\i} \Bigr]_0, \quad
       \Bigl[ \braketc{\f}{Q^{(-)}_{22}}{\bar \i} \Bigr]_0
     = \Bigl[ \braketc{\bar \f}{Q^{(-)}_{22}}{\i} \Bigr]_0,
           \eqno(\B.4\a)
$$
$$
       \Bigl[
        \frac{d\braketc{\bar \f}{Q^{(+)}_{21}}{\bar \i}}
          {d\omega_\rot} \Bigr]_0
     = \Bigl[
        \frac{d\braketc{\f}{Q^{(+)}_{21}}{\i}}
          {d\omega_\rot} \Bigr]_0 , \quad
       \Bigl[
        \frac{d\braketc{\f}{Q^{(-)}_{21}}{\bar \i}}
          {d\omega_\rot} \Bigr]_0
     = \Bigl[
        \frac{d\braketc{\bar \f}{Q^{(-)}_{21}}{\i}}
          {d\omega_\rot} \Bigr]_0 ,
           \eqno(\B.4\b)
$$
and in addition to them the identities between the matrix elements
of the operators with opposite signatures,
$$
       \Bigl[ \braketc{\bar \f}{Q^{(-)}_{22}}{\i} \Bigl]_0
  =\mp \Bigl[ \braketc{\f}{Q^{(+)}_{22}}{\i} \Bigl]_0 , \quad
       \Bigl[
        \frac{d\braketc{\bar \f}{Q^{(-)}_{21}}{\i}}
          {d\omega_\rot} \Bigr]_0
  =\mp \Bigl[
        \frac{d\braketc{\f}{Q^{(+)}_{21}}{\i}}
          {d\omega_\rot} \Bigr]_0 , \quad (\dK = \mp 2),
           \eqno(\B.4\c)
$$
can be derived.  These and similar identities for the $M1$ operators
were used to derive Eqs.(3.19) and (3.20) for the transitions
from the $\gamma$ band.

\vfill
\eject

%
%
\baselineskip 20 truept
\outrefs

\vfill
\eject

\end